\documentclass[12pt]{article}
\pdfoutput=1
\usepackage[square,comma,numbers,sort&compress]{natbib}
\usepackage{graphicx,epstopdf,amssymb,amsfonts,amsmath,amsthm,array,
mathrsfs,amscd}
\usepackage{float}
\usepackage{xcolor}
\usepackage{appendix}
\usepackage{hyperref}
\newcommand{\pbs}[1]{\let\temp=\\#1\let\\=\temp}
\numberwithin{equation}{section}
\def\be{\begin{equation}}\def\ee{\end{equation}}
\def\argtanh{\mathop{\text{argtanh}}\nolimits}

\def\cvp{\raise 2pt\hbox{,}} 
 \def\tr{\mathop{\text{tr}}\nolimits}

 \def\d{{\rm d}} 
\def\la{\lambda}

\def\slR{\text{SL}(2,\mathbb R)}\def\PslR{\text{PSL}(2,\mathbb R)}\def\aslR{\mathfrak{sl}(2,\mathbb R)}

\def\la{\lambda}

\def\mani{\mathscr M}\def\disk{\mathscr D}

\def\vphi{\varphi}

\def\uone{\text{U}(1)}
\def\plb#1#2#3{{\it Phys.\ Lett.\ }{\bf B #1} (#2) #3}
\def\npb#1#2#3{{\it Nucl.\ Phys.\ }{\bf B #1} (#2) #3}

\def\prl#1#2#3{{\it Phys.\ Rev.\ Lett.\ }{\bf #1} (#2) #3}
\def\jhep#1#2#3{{\it J. High Energy Phys.\ }{\bf #1} (#2) #3}
\def\prd#1#2#3{{\it Phys.\ Rev.\ }{\bf D #1} (#2) #3}

\def\cmp#1#2#3{{\it Comm.\ Math.\ Phys.\ }{\bf #1} (#2) #3}

\def\ijmpa#1#2#3{{\it Int.\ J.\ Mod.\ Phys.\ }{\bf A #1} (#2) #3}
\def\mpla#1#2#3{{\it Mod.\ Phys.\ Lett.\ }{\bf A #1} (#2) #3}

\def\imath#1#2#3{{\it Invent math }{\bf #1} (#2) #3}

\def\jpa#1#2#3{{\it J.\ Phys.\ }{\bf A #1} (#2) #3}

\begin{document}
%
%
{\pagestyle{empty}
\parskip 0in
\

\vfill
\begin{center}
{\LARGE\sffamily Gauge Theory Formulation of Hyperbolic Gravity}



\vspace{0.4in}


Frank F{\scshape errari}

\medskip
{\it Service de Physique Th\'eorique et Math\'ematique\\
Universit\'e Libre de Bruxelles (ULB) and International Solvay Institutes\\
Campus de la Plaine, CP 231, B-1050 Bruxelles, Belgique}

%

\smallskip
{\tt frank.ferrari@ulb.ac.be}
\end{center}
\vfill\noindent

We formulate the most general gravitational models with constant negative curvature (``hyperbolic gravity'') on an arbitrary orientable two-dimensional surface of genus $g$ with $b$ circle boundaries in terms of a $\PslR_{\partial}$ gauge theory of flat connections. This includes the usual JT gravity with Dirichlet boundary conditions for the dilaton field as a special case. A key ingredient is to realize that the correct gauge group  is not the full $\PslR$, but a subgroup $\PslR_{\partial}$ of gauge transformations that go to $\uone$ local rotations on the boundary. We find four possible classes of boundary conditions, with associated boundary terms, that can be applied to each boundary component independently. Class I has five inequivalent variants, corresponding to geodesic boundaries of fixed length, cusps, conical defects of fixed angle or large cylinder-shaped asymptotic regions with boundaries of fixed lengths and extrinsic curvatures one or greater than one. Class II precisely reproduces the usual JT gravity. In particular, the crucial extrinsic curvature boundary term of the usual second order formulation is automatically generated by the gauge theory boundary term. Class III is a more exotic possibility for which the integrated extrinsic curvature is fixed on the boundary. Class IV is the Legendre transform of class II; the constraint of fixed length is replaced by a boundary cosmological constant term.

\vfill

\medskip
%
\begin{flushleft}
\today
\end{flushleft}
\newpage\pagestyle{plain}
\baselineskip 16pt
\setcounter{footnote}{0}

}


\section{Introduction}

Since the founding papers of Polyakov \cite{Polyakov}, two-dimensional quantum gravity, which is the study of random two-dimensional geometries on a surface $\mani$, has remained a central subject in theoretical physics, with an impressively wide range of applications, from string theory and quantum gravity per se to statistical physics and probability theory. 

We focus on a particular class of theories in Euclidean signature for which the Ricci scalar $R$ is fixed to a constant negative value.\footnote{The Gaussian curvature, more commonly used by mathematicians, is $K=R/2$.} By choosing appropriately the units of length, we thus set
\be\label{Rcst} R = -2\, .\ee
Such metrics are called hyperbolic\footnote{In the mathematical literature, hyperbolic metrics must sometimes satisfy particular boundary conditions. For us, a hyperbolic metric is simply a metric for which $R=-2$.} and we name the associated quantum gravity theories, for which we sum over hyperbolic metrics only, \emph{hyperbolic gravity models}.

Hyperbolic gravity can be defined on an arbitrary oriented Riemann surface $\mani$ of genus $g$ and $b$ circle boundary components,
\be\label{bdform} \partial\mani = \bigcup_{i=1}^{b}\mathcal C_{i}\, .\ee
In the quantum gravity path integral, the constraint \eqref{Rcst} may be enforced by a Lagrange multiplier scalar ``dilaton'' field $\Phi$ in the gravitational action \cite{JTpapers},
\be\label{Sdil} S_{\text{grav.}} = -\frac{1}{16\pi}\int_{\mani}\!\d^{2}x\sqrt{g}\,\Phi\bigl(R +2\bigr) + \cdots\ee
where the dots indicate possible additional boundary terms when $\mani$ has a non-trivial boundary. 

Hyperbolic gravity turns out to have qualitatively very different incarnations, depending on the type of boundary conditions one imposes. 

The simplest case is when $\mani$ has no boundary, $b=0$. If the Euler characteristics $\chi(\mani)=2-2g-b$ is positive or zero, the Gauss-Bonnet formula implies that there is no $R=-2$ metric at all. When $\chi(\mani)<0$, the uniformization theorem implies that the space of $R=-2$ metrics is in one-to-one correspondence with the moduli space of complex structures on $\mani$, yielding a finite, $(6g-6)$-dimensional space over which to integrate. This can be fruitfully generalized by considering manifolds with geodesic boundaries of fixed lengths $(\ell_{1},\ldots,\ell_{b})$.\footnote{Manifolds with marked points are a special case for which the lengths $\ell_{i}$ go to zero.} The path integral then computes the volumes $V(\ell_{1},\ldots,\ell_{b})$ of the associated $(6g-6+2b)$-dimensional moduli space. This problem was brilliantly studied by Mirzakhani in \cite{Mirzakhani} by using recursion relations between the volumes, making the link with previous famous work by Witten and Kontsevitch \cite{Wittentopo}. See also \cite{DWreview} for a recent nice review on this subject. Further generalizations of the geodesic boundary conditions will appear naturally in the construction presented later in our paper.

The above models, as interesting as they are from the mathematical point of view, are far too simple to describe interesting physics. One needs, at the very least, an infinite dimensional space of metrics over which to integrate. One way to achieve this is to waive the constant curvature constraint. On top of the integral over moduli space, one then has to integrate over metrics in given conformal classes, yielding the famous Liouville theories of two-dimensional gravity, see e.g.\ \cite{2dgravcont}. But this is not what we are interested in: we want to stick with the constraint $R=-2$. The way to go is then to consider new boundary conditions. Instead of choosing the boundaries to be geodesics, one imposes Dirichlet boundary conditions on the dilaton field $\Phi$,
\be\label{PhiJTbc} \Phi = \vphi_{i} = \text{constant on }\mathcal C_{i}\, .\ee
The lengths $\ell_{i}$ of the boundaries remain fixed. The variational principle with these boundary conditions is made consistent by adding
\be\label{JTbdaction} S_{\text{JT}} = -\frac{1}{8\pi}\oint\!\d s\,\Phi k\ee
to the bulk action \eqref{Sdil}, where $k$ is the extrinsic curvature of the boundary and $\d s$ the length element. The term \eqref{JTbdaction} is added on each boundary component for which the condition \eqref{PhiJTbc} is imposed. We call \emph{JT gravity} this new version of hyperbolic gravity, for which the scalar field $\Phi$ plays a crucial role.\footnote{This is a convenient terminology, but one may as well call ``JT gravity'' any $R=-2$ model.} 

In a sense, the JT gravity models sits in between the ``topological gravity'' models studied by Witten, Kontsevitch, Mirzakhani and others, and the Liouville models: they have an infinite dimensional space of metrics and yet they retain the constant negative curvature condition. They turn out to have an extremely interesting physics, which has been intensely studied over the last few years. They provide holographic models \cite{holomod} for  strongly coupled quantum mechanical systems \cite{Kitaev,tensors} that are believed to describe the physics of near-extremal black holes. As such, they yield the first examples of the holographic correspondence for which the gravitational theory can be studied directly at the quantum level. Remarkable and puzzling properties have been discovered. For instance, in \cite{SSS}, combining the new ingredients of JT gravity with Mirzakhani insights and Eynard's topological recursion \cite{Eynard}, a beautiful relationship with matrix integrals was established, suggesting that the models may not be dual to genuine quantum systems but rather to random ensembles of such systems.

Theories of constant negative curvature metrics are related to theories of flat $\PslR=\text{SO}(2,1)_{+}$ connections \cite{pslform}. To see this, we use the algebra satisfied by the rotation and boost generators $J$, $K_{1}$ and $K_{2}$ of $\mathfrak{sl}(2,\mathbb R)=\mathfrak{so}(2,1)$,
\be\label{soalgebra} [J,K_{1}]=-K_{2}\, ,\quad [J,K_{2}]=K_{1}\, ,\quad [K_{1},K_{2}]=J\, .\ee
Expanding the gauge field
\be\label{Aexpansion} A = \omega J + e^{a}K_{a}\ee
and using \eqref{soalgebra} to compute the curvature two-forms, one finds
\be\label{Fform} F = \d A + A\wedge A = \bigl(\d\omega + \Omega\bigr) J + T^{a}K_{a}\, ,\ee
for
\begin{align}\label{volformdef}& \Omega = e^{1}\wedge e^{2}\\
\label{torsionformdef}&T^{a} = \d e^{a} + \epsilon^{a}_{\ b}\, \omega\wedge e^{b}\, .
\end{align}
The flatness condition $F=0$ is thus equivalent to
\begin{align}\label{flconn1} T^{a} & =0\\
\label{flconn2} \d\omega &= -\Omega\, .
\end{align}
Assuming that $\Omega\not = 0$, we can identify $e^{a}=e^{a}_{\mu}\d x^{\mu}$ with a zweibein and $\omega$ with a spin connection. We also have a metric and a volume form
\be\label{meteadef} g_{\mu\nu}=\delta_{ab}e^{a}_{\mu}e^{b}_{\nu}\, ,\quad \Omega_{\mu\nu}=\epsilon_{ab}e^{a}_{\mu}e^{b}_{\nu}\, .\ee
Equations \eqref{flconn1} and \eqref{flconn2} are then equivalent to vanishing torsion and constant negative curvature $R=-2$.

This beautiful correspondence suggests that hyperbolic gravity may be equivalent to a $\PslR$ gauge theory. Such a gauge theoretic formulation, reminiscent of the Chern-Simons formulation of three-dimensional gravity, would be of significant theoretical and practical interest. When $\mani$ is closed, the story is actually completely straightforward. One considers the so-called BF gauge theory action
\be\label{SBulk} S_{\text{Bulk}}= \frac{1}{4\pi}\int_{\mani}\tr \bigl(B F\bigr)\, ,\ee
where $B$ is a scalar field in the adjoint representation. We have conveniently normalized the action, with $\tr K_{1}^{2}=\tr K_{2}^{2} = -\tr J^{2} = 1/2$.\footnote{In most of the literature, the Euclidean action \eqref{SBulk} appears with an $i$ in front; we prefer to absorb the factor $i$ in the field $B$, consistently with the gravitational formulation in which the contour of integration for the Lagrange multiplier field $\Phi$ in \eqref{Sdil} is parallel to the imaginary axis.} The field $B$ plays the role of a Lagrange multiplier enforcing the flatness condition $F=0$. This simple action delivers the correct theory. However, in the presence of boundaries, which is by far the most interesting case, things are much more subtle.

The simplest way to understand the difficulty is to consider the case of JT gravity on a disk, $\mani=\disk$. On the one hand, the space of hyperbolic metrics on the disk is infinite dimensional. For instance, an important class of metrics, well-known in the physics literature, can be obtained from ``boundary reparameterization'' functions; in the limit $\ell\rightarrow\infty$, $\vphi\rightarrow\infty$ for fixed $\vphi/\ell$, the boundary term \eqref{JTbdaction} then yields the famous Schwarzian action \cite{holomod}. On the other hand, since the disk is simply connected, the space of flat connection is trivial. Any flat connection can be gauge-transformed to zero. A standard $\PslR$ theory of flat connections on the disk is thus uninteresting and cannot reproduce JT gravity.

Attemps have been made in the literature to cure this problem. For instance, in \cite{Verlinde19}, the authors propose to make the gauge theory on the disk non-trivial by adding by hand a loop defect in the bulk. The degrees of freedom then live on the defect. This is philosophically similar to the study of defects in \cite{MertensetalDefects}. These models can probably match with the gravitational description only \emph{approximately}. They were built to reproduce the physics in the case of the Schwarzian limit and are certainly useful for that purpose.

The goal of the present paper is to construct an exact gauge theoretic formulation of hyperbolic gravity. This formulation does not involve the addition of new degrees of freedom or defects on top of the gauge field. A special case of the theory is equivalent to the JT boundary conditions and generates automatically the correct extrinsic curvature term \eqref{JTbdaction} found in the gravitational picture. Our general formulation also accomodates the old topological gravity models and other interesting generalizations.

A central issue we have to deal with is that the gauge group $\PslR$ is too big. We need to break $\PslR$ by appropriate boundary conditions and/or boundary terms to generate more degrees of freedom. In particular, the correct theory on the disk must be non trivial.

Some simple ways to do this are well-known, but are not consistent with a gravitational interpretation. To clearly understand the problem, it is instructive to give an example. Let us consider the case of the disk. We pick a privileged parameter $p$ on the circle boundary, denote by $u$ the associated tangent vector and impose the boundary condition
\be\label{groupbc}  A(u)=\kappa B\quad\text{on $\partial\mani$,}\ee
where $\kappa$ is a coupling constant. The action 
\be\label{Sgroup} S = \frac{1}{4\pi}\biggl[\int_{\mani}\tr \bigl(B F\bigr)
-\frac{1}{2}\oint_{\partial\mani}\tr\bigl(B A\bigr)\biggr]\ee
has a well-defined variational principle for the b.c.\ \eqref{groupbc}. After integrating out $B$, the flatness condition yields $A=g^{-1}\d g$ for a globally defined map $g:\disk\rightarrow\PslR$. The boundary term, which reads
\be\label{groupparticle} -\frac{1}{8\pi\kappa}\oint_{\partial\disk}\!\d p \,\tr\Bigl(
g^{-1}\frac{\d g}{\d p}g^{-1}\frac{\d g}{\d p}\Bigr)\, ,\ee
describes the dynamics of a particle moving on the group manifold $\PslR$. This model is interesting, because it shows how a non-trivial dynamics can emerge from the breaking of $\PslR$ on the boundary, but it is not relevant for gravity. The reason is that it does not have the right symmetries: it breaks the boundary reparameterization symmetry and is invariant only under $\PslR$ gauge transformations that go to a constant on the boundary. The gauge symmetry is too small!

The correct gauge symmetries required to match with a gravitational model are easy to identify. We must preserve full diffeomorphism invariance, including reparameterization invariance on the boundary. We must also preserve the local $\uone$ subgroup of $\PslR$ that acts on the indices $a$ in \eqref{Aexpansion}, including on the boundary, because this subgroup is related to the local frame rotations in the gravitational picture. We call the gauge transformations that to go to these $\uone$ transformations on the boundary, $\PslR_{\partial}$ gauge transformations. Our strategy is to build the most general model with precisely these symmetries. We seek $\PslR_{\partial}$-invariant and diffeomorphism invariant boundary conditions and associated boundary terms which, added to the bulk action \eqref{SBulk}, yield a consistent variational principle.

Following this line of reasoning, we find four classes of possible boundary conditions. 

\noindent\textbf{Class I} amounts to fixing the length $\ell$ and the extrinsic curvature $k$ of the boundary. There are three qualitatively distinct cases to consider: $k<-1$, $-1<k<1$ and $k>1$; the cases $k=-1$ and $k=1$ can be understood as limits of the other three cases. The usual topological gravity boundary conditions \`a la Mirzakhani corresponds to $k=0$, but it happens that any choice $-1<k<1$ is equivalent to $k=0$. The case $k<-1$ turns out to be equivalent to inserting conical singularities of arbitrary angles, an instance that plays a role in recent studies of generalizations of the standard JT gravity \cite{Wittensummer}. The case of boundaries with $k>1$ is less familiar. Their existence puts strong constraints on the allowed topologies. For instance, one shows that if conical singularities with angles strictly greater than $2\pi$ are not present as well, the only possibility with $k>1$ boundaries is the disk.

\noindent\textbf{Class II} yields the usual JT gravity, with a boundary of fixed length and a boundary action \eqref{JTbdaction} given by the integrated extrinsic curvature. 

\noindent\textbf{Class III} amounts to fixing the integrated extrinsic curvature on the boundary. The boundary action is proportional to the length, i.e.\ we have a boundary cosmological constant.

\noindent\textbf{Class IV} is the Legendre transform of class II: the length is no longer fixed but a boundary cosmological constant is turned on, on top of the integrated extrinsic curvature term. 

The plan of the paper is as follows. In Section \ref{ReviewSec} we briefly review the usual first order formulation of gravity. We explain how this formulation is extended to introduce the $\PslR$ gauge symmetry. We write down the gauge theory bulk equations of motion and note that they match with the gravitational equations of motion of the second order formalism. We then address the crucial issue of the relationship between the $\PslR$ gauge symmetries and the gravitational gauge symmetries. In Section \ref{BcSec}, we discuss the possible boundary conditions that are consistent with $\PslR_{\partial}$ and diffeomorphism invariance. We construct the associated boundary terms and show that the variational principle is consistent. Finally, in Section \ref{GeomSec}, we interpret our four classes of boundary conditions in gravitational terms and discuss some of their basic properties, mainly focusing on class I.

The present paper is extracted from a set of lecture notes written by the author on various elementary aspects of hyperbolic gravitational theories \cite{Ferlectures}. These lecture notes are written in a pedagogical way and contain many more topics and details than the present manuscript. They will be published elsewhere.

\noindent\emph{Note}: when the present paper was in its final stage of preparation, an interesting work discussing the possible boundary conditions in the usual second order formulation of hyperbolic gravity, starting from the action \eqref{Sdil}, appeared \cite{Goel}. As far as we can see, the results of \cite{Goel} are nicely consistent with the  set of boundary conditions that we find from the gauge theory point of view; see in particular the discussion in Sec.\ \ref{gravinterSec}.

\section{\label{ReviewSec} Basic concepts}
\subsection{\label{firstorderSec} Basics of the first order formalism of gravity}

Let us consider an abstract $\text{SO(2)}=\text{U}(1)$ vector bundle $E$ over our surface $\mani$, with canonical Euclidean metric $\delta$ on the fibers, which is isomorphic to the tangent bundle $T\mani$. The degrees of freedom in the first order formalism are the $\text{U}(1)$ connections $D$ on $E$ and the isomorphisms between $E$ and $T\mani$.  

An isomorphism is parameterized locally by a zweibein matrix $(e^{a}_{\mu})$ or equivalently by a locally defined one-form with values in the space of sections of $E$,
\be\label{eadefAbs} e^{a} = e^{a}_{\mu}\,\d x^{\mu}\, ,\ee
such that $\Omega = e^{1}\wedge e^{2}$ is nowhere vanishing. Then, $(e^{1},e^{2})$ forms a basis and the dual basis is identified with an oriented orthonormal frame $(e_{1},e_{2})$. In other words, the isomorphism yields a Riemaniann metric and a volume form on $\mani$, given explicity by Eq.\ \eqref{meteadef}. Under the isomorphism, the abstract $\text{U}(1)$ gauge group of $E$ is  identified with the local orthogonal frame rotations on $T\mani$ and the $\text{U}(1)$ connection $D$ becomes a metric-preserving spin connection $\omega$ on $T\mani$, acting as
\be\label{Nablaabsact} D e_{a} = \epsilon^{b}_{\ a}\omega  \otimes e_{b}\, .\ee
The antisymmetric symbol $\epsilon$ is defined by $\epsilon^{a}_{\ b}=\epsilon_{a}^{\ b}=\epsilon_{ab}=\epsilon^{ab}=-\epsilon^{ba}$ and $\epsilon_{12}=+1$. Note that the torsion \eqref{torsionformdef} is a priori non-vanishing. Conversely, picking a metric, volume form and metric-preserving spin connection on $\mani$ yield a zweibein $(e^{a}_{\mu})$ and thus an isomorphism with $E$ together with an orientation and a $\text{U}(1)$ connection on $E$.

\subsection{\label{PSLbundleSec} $\PslR$ connections}

We now construct a $\PslR$ vector bundle $\mathscr E\rightarrow\mani$ from the $\text{U}(1)$ bundle $E\rightarrow\mani$ in the following way. Recall that $\PslR$ is isomorphic to the three-dimensional proper orthochronous Lorentz group $\text{SO}(2,1)_{+}$, the adjoint representation of $\PslR$ corresponding to the vector representation of $\text{SO}(2,1)_{+}$.

We choose the fibers of $\mathscr E$ to be $\mathbb R^{3}$, with sections $s=(s^{A})=(s^{0},s^{1},s^{2})$ transforming as Lorentz three-vectors. The topological structure of $\mathscr E$ is fixed by identifying the $\text{U}(1)$ gauge group of $E$ with the rotation subgroup of $\text{SO}(2,1)_{+}$ keeping $(1,0,0)$ fixed and by using this identification to set the transition functions of $\mathscr E$ consistently with those of $E$. 

\noindent\emph{Remarks}: 

\noindent i) For the important case of the disk, the situation is very simple since the bundles $E$ and $\mathscr E$ are trivial. However, the above construction is necessary in the general case to specify $\mathscr E$.

\noindent ii) The gauge group is $\PslR$ instead of $\slR$ because a $2\pi$ frame rotation corresponds to the identity.

\noindent iii) By construction, the bundle $\mathscr E\rightarrow\mani$ has a nowhere vanishing timelike section $s_{\text r}=(1,0,0)$ characterized by the fact that it is fixed under $\uone$. Conversely, if a $\PslR$ vector bundle admits such a section $s_{\text r}$, then we have a distinguished $\uone$ stabilizer subgroup of $\PslR$ that fixes $s_{\text r}$. The structure group can then be reduced to this $\text{U}(1)$, by choosing local trivializations for which the section is $(1,0,0)$. In our case, the resulting $\uone$ bundle is $E\rightarrow\mani$ and is isomorphic to $T\mani$.

Let us now pick a connection $\mathcal D$ on $\mathscr E$, with connection one-form expanded locally as in \eqref{Aexpansion}. Gauge transformations act as
\be\label{gaugetrans} A' = gAg^{-1} - (\d g) g^{-1}\ee
where $g\in\PslR$. For an infinitesimal transformation
\be\label{ginf} g = \mathbb I - \varepsilon = \mathbb I - \alpha J - \eta^{a}K_{a}\, ,\ee
we get
\be\label{delg1} \delta A = \mathcal D\varepsilon = \d\varepsilon + [A,\varepsilon]\ee
or equivalently
\begin{align}\label{delgomega} &\delta\omega = \d\alpha + \epsilon_{ab}e^{a}\eta^{b}\\
\label{delgea} &\delta e^{a} = \d\eta^{a}+ \epsilon^{a}_{\ b}\eta^{b}\omega - \alpha\epsilon^{a}_{\ b} e^{b}\, .\end{align}
The generator $J$ generates the rotation subgroup of $\PslR$. The transformation laws imply that $\omega$ yields a $\text{U}(1)$ connection on $E$ and that $(e^{1},e^{2})$ and $(\eta^{1},\eta^{2})$ transform as vectors under $\text{U}(1)=\text{SO}(2)$. Denoting by $D$ the $\text{U}(1)$ covariant derivative, we can rewrite \eqref{delgea} as
\be\label{delgea2} \delta e^{a} = D\eta^{a}- \alpha\epsilon^{a}_{\ b} e^{b}\, .\ee

We thus see that the degrees of freedom of the first order formulation of gravity, a $\text{U}(1)$ connection $\omega$ on $E$ and an isomorphism $e^{a}$ between $E$ and $T\mani$, are elegantly encoded in a $\PslR$ connection. Conversely, a $\PslR$ connection \eqref{Aexpansion} such that $\Omega=e^{1}\wedge e^{2}\not = 0$ yields an isomorphism $e^{a}$ with the tangent bundle and a spin connection $\omega$. 

Moreover, as already reviewed in the introduction section, flat connections satisfy \eqref{flconn1} and \eqref{flconn2}. The $\uone$ connection $D$
%
%
is then the Levi-Civita connection and the curvature is $R=-2$. The conclusion is that \emph{constant negative curvature metrics on $\mani$ are in one-to-one correspondence with flat $\PslR$ connections for which $\Omega\not =0$.} 

\noindent\emph{Remarks on the condition $\Omega\not = 0$}:

\noindent i) Clearly, the gauge theoretic and the metric formulations can be equivalent only if we discard the gauge field configurations for which $\Omega\not = 0$. This is familiar. This condition is manifestly invariant under $\uone$ gauge transformations but not under general $\PslR$ gauge transformations. This important subtlety will be discussed below in Sec.\ \ref{bcgaugeSec}.

\noindent ii) It is natural to wonder whether the inclusion of the $\Omega=0$ configurations is physically relevant or might even be required (or forbidden) for non-perturbative consistency. The idea is that singular space-times might have to be added in the gravitational path integral. We shall not discuss this possibility here. Our point of view is that, at least in two dimensions, the metric formulation of gravity is most probably consistent, even non-perturbatively, as long as we fix the topology of $\mani$.\footnote{The situation in three dimensions may be subtler. We also do not discuss the summation over all possible topologies here.}

\subsection{\label{BulkSecSec}Bulk action and bulk equations of motion}

The bulk action \eqref{SBulk} is manifestly invariant under $\PslR$ gauge transformations, under which $B$ and $F$ transform as
\be\label{delgchiz}\delta B = [B,\varepsilon]\, ,\quad \delta F = [F,\varepsilon] .\ee
Expanding $B$ as
\be\label{chiexp} B = \Phi J + \chi^{a}K_{a}\, ,\ee
this is equivalent to
\begin{align}\label{delgPhi} \delta\Phi &= \epsilon_{ab}\chi^{a}\eta^{b}\\
\label{delgchi}\delta\chi^{a} &= \epsilon^{a}_{\ b}\eta^{b}\Phi -\alpha\epsilon^{a}_{\ b}\chi^{b}\
\end{align}
and similar variation formulas for the components of $F$ in \eqref{Fform}. The action is also invariant under the usual space-time gauge symmetries, diffeomorphisms and local frame rotations. This is so even though it does not depend on a choice of metric on $\mani$. We say that the action is topological.  

Varying the gauge field $A$ and Lagrange multiplier field $B$ in \eqref{SBulk} yield the bulk equations of motion
\be\label{gaugeeom} F=0\, ,\quad \mathcal D B = \d B + [A,B] = 0\, .\ee
Using the expansions \eqref{Fform} and \eqref{chiexp}, this may be written in components as \eqref{flconn1}, \eqref{flconn2} and 
\begin{align}\label{Phigeom}& \d\Phi + \epsilon_{ab}e^{a}\chi^{b} = 0\\
\label{chigeom} & D\chi^{a}-\Phi\epsilon^{a}_{\ b}e^{b}= \d\chi^{a}+\epsilon^{a}_{\ b}\chi^{b}\omega-\Phi\epsilon^{a}_{\ b}e^{b}=0\, .
\end{align}
It can be straightforwardly checked that these equations are equivalent to the gravitational equations of motion obtained by varying the action \eqref{Sdil}, the gauge theory scalar $\Phi$ and the dilaton field $\Phi$ being identified. We also find that
\be\label{chimuclass} \chi^{\mu}=e^{\mu}_{a}\chi^{a}=\Omega^{\mu\nu}\partial_{\nu}\Phi\ee
is a Killing vector.

For future reference, let us write down the consequences of the equations of motion on the boundary. Applying \eqref{Phigeom} and \eqref{chigeom} on the unit tangent and normal vectors (and/or using \eqref{chimuclass}), we get
\begin{align}\label{eomonbd1} &\chi^{a} = -n^{\mu}\partial_{\mu}\Phi\, t^{a} + \frac{\d\Phi}{\d s}\, n^{a}\\\label{eomonbd2} &D_{t}\chi^{a} = \Phi\, n^{a}\, ,\quad D_{n}\chi^{a} = -\Phi\, t^{a}\, .
\end{align}

\subsection{\label{bcgaugeSec}Gauge symmetries and diffeomorphisms}

The first order gauge theory description involves a $\PslR$ gauge symmetry. The second order gravitational description involves diffeomorphisms and local rotations gauge symmetries. We have already identified the local rotations with the $\text{U}(1)$ subgroup of $\PslR$. What about the diffeomorphisms?

Diffeomorphisms generated by a vector field $\xi$ and local frame rotations of infinitesimal angle $\vartheta$ act as
\begin{align}\label{deldiffomega} \delta\omega &= \mathcal L_{\xi}\omega + \d\vartheta\\
\label{deldiffea} \delta e^{a} &=\mathcal L_{\xi}e^{a}- \vartheta\epsilon^{a}_{\ b} e^{b}\\
\label{deldiffPhi} \delta\Phi &= \xi^{\mu}\partial_{\mu}\Phi\\
\label{deldiffchi} \delta\chi^{a} &= \xi^{\mu}\partial_{\mu}\chi^{a} -\vartheta\epsilon^{a}_{\ b}\chi^{b}\, ,
\end{align}
where $\mathcal L$ is the usual Lie derivative. In general, these transformations are different from $\PslR$ gauge transformations. However, when the bulk equations of motion \eqref{gaugeeom} are imposed, it is straightforward to check that \eqref{delgomega}, \eqref{delgea}, \eqref{delgPhi}, \eqref{delgchi} and \eqref{deldiffomega}, \eqref{deldiffea}, \eqref{deldiffPhi}, \eqref{deldiffchi} formally coincide, with a correspondence between the $\aslR$ and $\mathfrak{u}(1)\oplus\mathfrak{diff}(\mani)$ parameters given by
\be\label{paramid} \alpha = \vartheta + \omega(\xi)\, ,\quad \eta^{a}=e^{a}(\xi)\, .\ee
The relation between $(\alpha,\eta^{a})$ and $(\vartheta,\xi)$ is invertible if and only if $\Omega\not = 0$, as expected.

\emph{At this stage, a crucial difference occurs between the cases of closed manifolds and the cases of manifolds with boundary.}

\noindent\emph{Case one:} $\mani$ has no boundary. There is then an \emph{exact} identification between the $\PslR$ and the $\text{Diff}(\mani)\times\text{U}(1)$ gauge transformations when acting on flat $\PslR$ connections such that $\Omega\not = 0$. In particular, the condition $\Omega\not = 0$, being manifestly $\text{Diff}(\mani)\times\text{U}(1)$ invariant, is automatically $\PslR$ invariant as well. The gauge theoretic and the gravitational points of view agree.

\noindent\emph{Case two:} $\mani$ has a non-trivial boundary $\partial\mani$. There is then a mismatch between the $\PslR$ and the $\text{Diff}(\mani)\times\text{U}(1)$ gauge transformations, even in the case of the flat connections. This is so because \emph{the identification \eqref{paramid} does not respect the correct boundary conditions one must impose on the elements of $\mathfrak{diff}(\mani)$}. 

Indeed, in the standard gauge theory formulation, the infinitesimal gauge parameters $\alpha$ and $\eta^{a}$ are arbitrary smooth local functions. In particular, via the second equation in \eqref{paramid}, the parameter $\eta^{a}$ is associated to an \emph{arbitrary} vector field $\xi^{\mu}=e^{\mu}_{a}\eta^{a}$. However, in the gravitational theory, diffeomorphisms are not generated by arbitrary vector fields, but by \emph{vector fields that must be tangent to the boundary on the boundary.} Infinitesimal diffeomorphisms of $\mani$ restrict to infinitesimal diffeomorphisms on each connected component of $\partial\mani$. 

Note that no subtlety arises for local frame rotations, that are generated by an arbitrary smooth local function $\vartheta$ and can thus be identified with the $\text{U}(1)$ subgroup of $\PslR$ generated by $J$, even in the presence of a boundary.

We are here stumbling on a simple but fundamental issue in the gauge theory formulation, associated with the presence of boundaries. This issue is the core of the difficulty encountered to formulate correctly hyperbolic gravity in terms of a gauge theory. 

Let us define $\PslR_{\partial}$ gauge transformations to be $\PslR$ transformations that go to $\uone$ transformations on the boundary. More formally, these are gauge transformations that stabilize the nowhere vanishing timelike section $s_{\text r}$ of $\mathscr E\rightarrow\mani$ on the boundary. We are now going to build the most general boundary conditions and boundary terms that respect full diffeomorphism invariance, including reparameterization invariance on the boundary, together with $\PslR_{\partial}$.  These are precisely the gauge symmetries for which a gravitational interpretation of the theory will be valid. In particular, the off-shell $\PslR$ gauge invariance must be broken down to $\PslR_{\partial}$.

\section{\label{BcSec}Symmetries and boundary conditions}

To guide our discussion, a key point is that, when $\partial\mani$ is non-empty, the variation of the bulk action \eqref{SBulk} has a boundary term,
\be\label{Sgaugevar} \frac{1}{4\pi}\delta\int_{\mani}\tr\bigl(B F\bigr) = \frac{1}{4\pi}\int_{\mani}\tr\bigl(\delta B\, F + \delta A\wedge\mathcal D B\bigr) + \frac{1}{4\pi}\oint_{\partial\mani}\tr\bigl(B\delta A\bigr)\, .\ee
In order to make manifest the $\text{U}(1)$ symmetry transformation properties, we rewrite the boundary variation as
\be\label{Svarbduone} \frac{1}{4\pi}\oint\tr\bigl(B\delta A\bigr) = \frac{1}{8\pi}\oint\bigl(-\Phi\delta\omega + \delta_{ab}\chi^{a}\delta e^{b}\bigr)\, .\ee
If $u$ is a tangent vector to $\partial\mani$, the formula \eqref{Svarbduone} suggests to impose Dirichlet boundary conditions on $\omega(u)$ and $e^{a}(u)$ along $\mathcal C$. More precisely, taking into account our symmetry requirements, we could consider fixing $\omega(u)$ and $e^{a}(u)$ \emph{modulo $\text{U}(1)$ gauge transformations and reparameterizations.}

It is important to understand that fixing both $\omega(u)$ and $e^{a}(u)$ modulo $\text{U}(1)$ gauge transformations and reparameterizations is not the same as fixing $A(u)$ modulo $\PslR$ gauge transformations and reparameterizations. The latter would be equivalent to fixing the conjugacy class of the holonomy of the gauge field around the boundary, which preserves the full $\PslR$ gauge symmetry. The former breaks $\PslR$ down to $\PslR_{\partial}$. The difference between the two cases will be made completely explicit in Sec.\ \ref{bcIfullSec}.

Other choices of similar boundary conditions are possible. By adding appropriate diffeomorphism and $\PslR_{\partial}$ invariant boundary terms to the action, that will be written down explicity in Sec.\ \ref{corractionSec}, we can ``Legendre transform'' the boundary terms in \eqref{Svarbduone}, interchanging $-\Phi\delta\omega$ with $+\omega\delta\Phi$ and/or $\delta_{ab}\chi^{a}\delta e^{b}$ with $-\delta_{ab}\delta\chi^{a} e^{b}$. This yields four possible classes of boundary conditions, that we are now going to examine more closely.

We focus on a particular component $\mathcal C$ of $\partial\mani$. The circle $\mathcal C$ has an orientation derived from the orientation of $\mani$. We use an arbitrary parameter $p$ along $\mathcal C$, with associated tangent vector $u$, whose sign is always chosen consistently with the orientation of $\mathcal C$. As a preliminary, we first discuss the consequences of fixing $e^{a}(u)$ alone.

\subsection{Fixing $e^{a}(u) = u^{a}$}

Fixing $u^{a}$ modulo $\text{U}(1)$ gauge transformations and reparameterization is equivalent to the statement that the most general allowed infinitesimal variation of $u^{a}$ is of the form
\be\label{deltaeab} \delta u^{a} = \zeta\frac{\d u^{a}}{\d p} + \frac{\d\zeta}{\d p} u^{a} - \alpha\epsilon^{a}_{\ b}u^{b}\ee
for arbitrary well-defined (periodic along $\mathcal C$) functions $\zeta(p)$ and $\alpha(p)$. 

Let us assume first that the boundary condition is chosen such that $u^{a}$ is non-vanishing along $\mathcal C$. Note that this is a $\text{U}(1)$ and reparameterization invariant statement. This particular choice is particularly relevant if one wants to make the link with a gravitational theory, as will become clear shortly. The consequences of allowing $u^{a}$ to vanish will be discussed below. 

Picking $u^{a}\not = 0$ has two consequences that will come in handy later.

i) First, if we define 
\be\label{gaugelength} h = \delta_{ab}u^{a}u^{b}\ee
and 
\be\label{ellgauge} \ell = \int_{\mathcal C}\!\sqrt{h}\,\d p\, ,\ee
then we can show that imposing the boundary condition \eqref{deltaeab} is \emph{equivalent} to fixing $\ell$. One way of the equivalence is trivial: $\ell$ is expressed in terms of $u^{a}$ only and is manifestly $\text{U}(1)$ and reparameterization invariant. It is thus fixed by our boundary condition and corresponds to a parameter in the theory.\footnote{This is true even if we allow $u^{a}$ to vanish.} Conversely, if $\ell$ is fixed, the most general variation of $\smash{\sqrt{h}}$ must be the total derivative of a well-defined function on the boundary, $\smash{\delta\sqrt{h} = \d\tilde\zeta/\d p}$. Setting $\smash{\zeta = \tilde\zeta/\sqrt{h}}$ (this is where we use that $h$ is non-vanishing) then yields
\be\label{deltagbgauge}\delta h = \zeta \frac{\d h}{\d p} + 2 \frac{\d\zeta}{\d p}h\, ,\ee
which, using \eqref{gaugelength}, is equivalent to \eqref{deltaeab}.

Clearly, the condition $u^{a}\not = 0$, or equivalently $h>0$, allows to interpret
\be\label{dsgaugeth} \d s = \sqrt{h}\,\d p\ee
as a length element along $\mathcal C$. In particular, the transformation law of $h$ is the same as the one for the induced metric in the gravitational theory. This is so even if we do not have a metric on $\mani$ at this stage; in particular, $\Omega = e^{1}\wedge e^{2}$ may vanish both in the bulk and on the boundary. The point is that fixing $e^{a}(u)$ is enough to endow the boundary with a length element. Of course, the parameter $\ell$ will be matched to the length of the boundary component in the gravity picture.

ii) Second, a non-vanishing $u^{a}$ offers a privileged trivialization, or \emph{framing}, of the restriction of the $\text{U}(1)$ bundle $E$ over $\mathcal C$. 

A $\text{U}(1)$ bundle over a circle is always trivial, but there is no canonical framing. Two framings are related by a $\text{U}(1)$ gauge transformation, which is a map from $\mathcal C = \text{S}^{1}$ to $\uone=\text{S}^{1}$. If the winding number of this map is non-zero, the framings are said to be inequivalent.

Given $u^{a}$, the canonically associated framing can be defined as being the gauge in which $u^{2}=0$. This is well-defined only because $u^{a}$ is non-vanishing. Once this gauge is chosen, we can consider the $\uone$ connection $\bar\omega$ on $\mathcal C$ which is zero in this gauge. Explicitly, if we define the unit section
\be\label{tadef} t^{a}= \frac{u^{a}}{\sqrt{h}}\,\cvp\ee
then, in an arbitrary gauge,
\be\label{omebardef}\bar\omega = \epsilon_{ab}t^{a}\frac{\d t^{b}}{\d p}\,\d p\, .\ee
In terms of the ``angle'' $\gamma$ defined by $t^{1}+it^{2}=e^{i\gamma}$,\footnote{The map $e^{i\gamma(p)}$ is a global section of the principal $\text{U}(1)$ bundle on $\mathcal C$ and provides another equivalent way to define the privileged framing.} 
\be\label{ombar2} \bar\omega =\frac{\d\gamma}{\d p}\,\d p\, . \ee
In the gauge $u^{2}=0$, $\gamma=0$ and thus $\bar\omega = 0$, as announced. 

Let us summarize: fixing $u^{a}$ in such a way that $u^{a}$ is non-vanishing along $\mathcal C$ is equivalent to fixing the ``length'' \eqref{ellgauge}. It also provides a canonical framing of the $\uone$ bundle over $\mathcal C$. 

\noindent\emph{Remark}: if we wish to allow zeros of $u^{a}$, then $\ell$ does not contain all the boundary data. For instance, if we allow isolated zeros of $u^{a}$, then the theory also depends on the number of such zeros. We shall see in Sec.\ \ref{corractionSec} that considering this generalized, rather exotic situation (it is exotic in the sense that it does not have a standard gravitational counterpart) has rather drastic consequences when the theory is formulated on general surfaces $\mani$.

We can now discuss the four allowed classes of boundary conditions.

\subsection{The four classes of boundary conditions}

\subsubsection{\label{bcIfullSec}Boundary condition of class I} 

\noindent\textbf{B.c.\ I:} we fix $e^{a}(u) = u^{a}$ and $\omega(u)$ modulo $\text{U}(1)$ gauge transformations and reparameterization.

The consequences of fixing $u^{a}$ has been discussed above. Fixing $\omega(u)$ modulo $\text{U}(1)$ gauge transformations and reparameterization is equivalent to allowing
\be\label{deltaomeu} \delta\omega (u) = \zeta \frac{\d\omega(u)}{\d p} + \frac{\d\zeta}{\d p} \omega(u) + \d\alpha (u)\ee
on the boundary. This is equivalent to fixing the holonomy $\exp (i\oint_{\mathcal C}\omega)$ along $\mathcal C$, or equivalently the contour integral $\oint_{\mathcal C}\omega$ modulo $2\pi\mathbb Z$.

However, \emph{fixing both $u^{a}$ and $\omega(u)$ provides more information than fixing them independently.} Assuming that $u^{a}$ is non-vanishing, we can use the unit section \eqref{tadef} and the associated framing to build
\be\label{Komega} k_{\omega} = \bar\omega(t) - \omega(t)\, .\ee
This quantity is both $\uone$ and reparameterization invariant. It is thus fixed by our choice of boundary condition and constitutes a new parameter in the model. One may consider arbitrary profiles $k_{\omega}(s)$, or simply constant values $k_{\omega}$, as we shall do in the following. It is straightforward to check that fixing both $u^{a}$ and $\omega(u)$ is equivalent to fixing $\ell$ and $k_{\omega}$. 

To anticipate the equivalence with a gravitational picture, let us show that the quantity $k_{\omega}$ has a natural geometric interpretation. By using the non-vanishing mapping $u^{\mu}\mapsto u^{a}= e^{a}_{\mu}u^{\mu}$ from $T\mathcal C$ to the restriction of $E$ over $\mathcal C$, we can associate to the boundary curve $\mathcal C$ the quantity $D_{t}t^{a}$, where $t^{a}$ is defined in \eqref{tadef}. We claim that
\be\label{Dttuone} D_{t}t^{a} = -k_{\omega}\epsilon^{a}_{\ b}t^{b}\, .\ee
Indeed, since the $\uone$ connection preserves the canonical scalar product $\delta$ on the fibers of $E$, $D_{t}t^{a}$ must be orthogonal to $t^{a}$ and thus proportional to $\epsilon^{a}_{\ b}t^{b}$. Expanding $D_{t}t^{a} = \d t^{a}/\d s + \epsilon^{a}_{\ b}\omega(t) t^{b}$ then yields the coefficient of proportionality $\epsilon_{ab}t^{a}\d t^{b}/\d s - \omega(t)=\bar\omega(t) -\omega(t)=k_{\omega}$. We could call $k_{\omega}$ the extrinsic curvature of the boundary curve $\mathcal C$ with respect to $\omega$, given the mapping $u^{\mu}\mapsto u^{a}$.

To summarize, fixing $u^{a}$ and $\omega(t)$ on $\mathcal C$ is equivalent to fixing the ``length'' $\ell$ and the ``extrinsic curvature'' $k_{\omega} = k$. We get two parameters $(\ell, k)$ for each boundary component on which we apply this boundary condition.

\subsubsection{Boundary condition of class II} 

\noindent\textbf{B.c.\ II:} we fix $e^{a}(u) = u^{a}$ and $\Phi$ modulo $\text{U}(1)$ gauge transformations and reparameterization. 

Assuming that $u^{a}$ is non-vanishing, the above discussion shows that this is equivalent to fixing $\ell$ and $\Phi=\varphi$ on the boundary. We get two parameters $(\ell,\varphi)$ for each boundary components on which we apply this boundary condition. This is clearly reminiscent of the standard JT boundary condition in the gravity formulation.

\subsubsection{Boundary condition of class III} 

A first natural option to define the b.c.\ III is to fix $\chi^{a}$ and $\omega(u)$ modulo $\text{U}(1)$ gauge transformations and reparameterization. Fixing $\chi^{a}$ is equivalent to fixing its norm $\delta_{ab}\chi^{a}\chi^{b}(p)$ modulo reparameterization. A natural choice is then to pick $\sqrt{\delta_{ab}\chi^{a}\chi^{b}}$ to be a constant. However, this choice implies that $\delta_{ab}\chi^{a}D_{t}\chi^{b}=0$ on the boundary. Eq.\ \eqref{eomonbd1} and \eqref{eomonbd2} then yield
\be\label{onshellbdchi} \Phi = \text{constant}\, ,\quad \chi^{a} = -n^{\mu}\partial_{\mu}\Phi\, t^{a}\quad\text{on shell.}\ee
This indicates that $\chi^{a}$ must be tied-up to the unit tangent vector; but this will not follow from simply fixing $\chi^{a}$.

Taking into account the above remarks, we thus define

\noindent\textbf{B.c.\ III:} we fix $\omega(u)$ modulo $\text{U}(1)$ gauge transformations and reparameterization and we set 
\be\label{bcIIIset} \chi^{a} = -\la t^{a}\ee
for a constant coupling $\la$.

The most general variation of $\omega(u)$ on the boundary is thus as in \eqref{deltaomeu}. The condition \eqref{bcIIIset} assumes that $u^{a}$ is non-vanishing on the boundary, so that $t^{a}$ can be defined; this must be understood as being part of our boundary condition. The minus sign in \eqref{bcIIIset} is a convenient convention, see Sec.\ \ref{gravinterSec}. 
 
The non-vanishing of $u^{a}$ allows to define a privileged framing and the associated connection $\bar\omega$ as in \eqref{omebardef}, but the extrinsic curvature $k_{\omega}$ \eqref{Komega} is not fixed since $u^{a}$ itself is not. However, the integrated extrinsic curvature
\be\label{ointframing} \Upsilon= \oint_{\mathcal C}\bigl(\bar\omega-\omega\bigr) = \oint_{\mathcal C}\d s\, k_{\omega}\ee
is fixed. Indeed, it is manifestly $\uone$ and reparameterization invariant; its variation is given by the variation of $-\oint\omega$, since $\oint\bar\omega$ is an integer, and this is zero given the b.c.\ \eqref{deltaomeu} imposed on $\omega(u)$.

Overall, we thus get two parameters $(\la,\Upsilon)$ for each boundary components on which we apply this boundary condition.

\subsubsection{Boundary condition of class IV} 

\noindent\textbf{B.c.\ IV:} we fix $\Phi=\vphi$ and impose \eqref{bcIIIset} on $\chi^{a}$.

This ensures as for b.c.\ III that we have a privileged framing on the boundary. We get two parameters $(\la,\vphi)$ for each boundary components on which we apply this boundary condition.

\subsection{\label{corractionSec} Boundary terms and the variational principle}

\subsubsection{Construction of the boundary terms}

The action of the theory is 
\be\label{Sgtdef}S =S_{\text{Bulk}}+ S_{\text{Boundary}} = \frac{1}{4\pi}\int_{\mani}\tr \bigl( B F\bigr)
+ S_{\text{Boundary}}\, ,\ee
where the reparameterization and $\text{U}(1)$ invariant boundary term $S_{\text{Boundary}}$ picks contributions from each circle component of $\partial\mani$ on which we impose boundary conditions of class II, III or IV (no boundary term is needed for the b.c.\ I). Explicitly, focusing on a given circle component $\mathcal C$ at a time, we add the following terms.

For the b.c.\ II, assuming we choose $u^{a}$ to be non-vanishing, the boundary term is
\be\label{STwo} S_{\text{II}} = \frac{1}{8\pi}\oint_{\mathcal C}\Phi\bigl(\omega - \bar\omega\bigr)
=\frac{\varphi}{8\pi}\oint_{\mathcal C}\bigl(\omega - \bar\omega\bigr)
 \, .\ee
The term $\oint\bar\omega$ is just a constant, but its addition is crucially required to get a $\uone$ invariant action. The consequences of waiving the condition $u^{a}\not = 0$ will be discussed below. 

For the b.c.\ III, we add
\be\label{SThree} S_{\text{III}} = -\frac{1}{8\pi}\oint_{\mathcal C}\delta_{ab}\chi^{a} e^{b}\, .\ee
The integrand is manifestly $\text{U}(1)$ invariant. 

For the b.c.\ IV we add
\be\label{SFour} S_{\text{IV}}= \frac{1}{8\pi}\oint_{\mathcal C}\Bigl[\Phi\bigl(\omega-\bar\omega\bigr) -\delta_{ab}\chi^{a} e^{b}\Bigr] = \frac{\varphi}{8\pi}\oint_{\mathcal C}\bigl(\omega-\bar\omega\bigr)-\frac{1}{8\pi}\oint_{\mathcal C}\delta_{ab}\chi^{a} e^{b}\, .\ee
As for the b.c.\ II, the addition of the constant term $\oint\bar\omega$ is required to ensure full $\text{U}(1)$ invariance.

\subsubsection{Variational principle}

We claim that with the above boundary terms in the action, the variational principle is well-defined. In particular, all the boundary terms appearing in the variation of the action automatically vanish by using the bulk equations of motion only. 

For instance, for b.c.\ of type I, the boundary term in the variation of the action is \eqref{Svarbduone}. Taking into account \eqref{deltaeab} and \eqref{deltaomeu} and integrating by part, we get
\be\label{vartypeI} \delta S = \oint_{\mathcal C}\d p\biggl[\Bigl(\omega(u)\frac{\d\Phi}{\d p} - \delta_{ab}u^{a}\frac{\d\chi^{a}}{\d p}\Bigr)\zeta + \Bigl(\frac{\d\Phi}{\d p}-\epsilon_{ab}\chi^{a}u^{b}\Bigr)\alpha\biggr] + \cdots\, ,\ee
where the dots represent possible contributions from other boundary components. The term proportional to $\alpha$ vanishes from the bulk equation of motion \eqref{Phigeom}. The term proportional to $\zeta$ then also vanishes by contracting \eqref{chigeom} with $u^{a}$ and using again \eqref{Phigeom}.

For b.c.\ of type II, the boundary term in the variation of the action reads
\be\label{vartypeII1}\delta S = \frac{1}{8\pi}\oint_{\mathcal C}\bigl(\omega\delta\Phi + \delta_{ab}\chi^{a}\delta e^{b}\bigr)+ \cdots\, .\ee
Taking into account $\Phi = \text{constant}$ and \eqref{deltaeab} and integrating by part, we get
\be\label{vartypeII2} \delta S = -\frac{1}{8\pi}\oint_{\mathcal C}\d p\Bigl(\delta_{ab}u^{a}\frac{\d\chi^{a}}{\d p}\,\zeta + \epsilon_{ab}\chi^{a}u^{b}\alpha\Bigr)\, .\ee
This vanishes by using \eqref{Phigeom}, \eqref{chigeom} and the condition $\Phi=\text{constant}$.

We let the reader check the consistency of b.c.\ III and IV, along the same lines.

\subsubsection{Waiving the condition of non-vanishing $u^{a}$}

One may contemplate the possibility of waiving the condition of non-vanishing of $u^{a}$ for b.c.\ I and II. The condition \eqref{bcIIIset} for b.c.\ III and IV implicitly assumes that $u^{a}$ is non-vanishing, but we may waive this condition in the special case $\la=0$. In case I, the choice of b.c.\ can no longer be described in terms of fixing the length and the extrinsic curvature; in case III, since we cannot use $\bar\omega$ anymore, the $\mathbb R$-valued parameter $\Upsilon$ has to be replaced by $\oint\omega$, which is defined only modulo $2\pi\mathbb Z$. This being said, there is no inconsistency.

The situation is different for b.c.\ II and IV. To write the boundary terms \eqref{STwo} or \eqref{SFour} in the action, we crucially need a canonical framing. If we don't have it, the best we can do is to replace $\oint(\omega-\bar\omega)$ by $\oint\omega$. But such a term is not gauge invariant under $\text{U}(1)$ gauge transformation that have a non-zero winding number along the boundary. This is a \emph{framing anomaly}. This problem may be dealt with in two different ways.

We may decide to live with the anomaly, i.e.\ accept that the observables of our theory depend explicitly on a choice of framing. If $E$ or equivalently $T\mani$ is trivial, it is natural to work with the trivialization $E = \mani\times\mathbb R^{2}$ and there is thus a natural framing. This happens for instance in the case of the disk $\mani=\disk$. This is equivalent to defining $\oint\omega = \int\d\omega$, using the fact that $\omega$ is globally defined. In this point of view, the allowed $\uone$ gauge transformations must be globally defined on $\disk$ and such gauge transformations automatically have zero winding number on $\partial\disk$. The term $\oint\omega$ is thus well-defined in this restricted sense.

On the other hand, if we want to get rid on the framing anomaly, we need to impose the quantization condition
\be\label{phiquantization} \varphi\in 8i\pi\mathbb Z\ee
on each relevant boundary component. The quantization condition \eqref{phiquantization} is the analogue of level quantization in Chern-Simons theory. In the present context, this quantization condition may be interpreted as being an additional constraint required to make sense of certain singularities in the gravity picture, for instance when we allow the ``length element'' \eqref{dsgaugeth} to degenerate.

\section{\label{GeomSec} Discussion}

\subsection{\label{gravinterSec}Gravitational interpretation of the boundary conditions}

In order to get a standard gravitational interpretation, we restrict ourselves to the gauge field configurations for which $\Omega\not = 0$, see the discussion at the end of Sec.\ \ref{PSLbundleSec}. After the Lagrange multiplier field $B$ is integrated out, we get hyperbolic gravitational theories with various boundary conditions and boundary contributions that govern the dynamics of the model. 

\paragraph{Class I} The geometrical interpretation is straightforward. The parameters $\ell$ and $k_{\omega}$ defined by \eqref{ellgauge} and \eqref{Komega} match with the length and extrinsic curvature of the boundary. From the gravity point of view, we thus fix the lengths and extrinsic curvatures $(\ell_{i},k_{i})$ of the boundary components $\mathcal C_{i}$ on which the b.c.\ I are applied. There is no boundary action. This is an interesting generalization of the familiar set-up, considered by Mirzakhani, for which the boundaries are geodesics, $k_{i}=0$. We provide a more detailed discussion of this case in the next subsection.

\paragraph{Class II} As for b.c.\ I, after integrating out $B$, the length $\ell$ of the boundary is fixed and $k_{\omega}=\bar\omega(t)-\omega(t)$ matches with the extrinsic curvature. The boundary term \eqref{STwo} thus precisely reproduces the crucial extrinsic curvature boundary term \eqref{JTbdaction} found in the usual second order formulation of JT. We have thus succeeded in constructing a gauge theory version of this theory. By breaking the gauge symmetry down to $\PslR_{\partial}$, the correct degrees of freedom and boundary term have been generated to ensure the equivalence with the standard gravitational formulation.

\paragraph{Class III} In this case, the integrated extrinsic curvature $\oint\d s k$ is fixed and the boundary action \eqref{SThree} takes the form
\be\label{Sthree2} S_{\text{III}} =-\frac{1}{8\pi}\oint_{\mathcal C}\d s\, \delta_{ab}\chi^{a}t^{b}=\frac{\la}{8\pi}\,\ell\, .\ee
The parameter $\la$ thus plays the role of an effective boundary cosmological constant. To get a stable model we need to choose $\la>0$. 

On the disk, this model is particularly simple. By Gauss-Bonnet, fixing $\Upsilon=\oint\d s k$ is equivalent to fixing the total area $A$. We thus have a model for fixed area metrics with an action proportional to the boundary length. This is in some sense ``dual'' to the usual JT gravity, for which we fix the length of the boundary and for which the action \eqref{STwo},
\be\label{Stwo2} S_{\text{II}}= -\frac{\vphi}{8\pi}\oint_{\disk}\d s\, k= -\frac{\vphi}{4}-\frac{\vphi}{8\pi}\, A\, ,\ee
is, up to a trivial constant, proportional to the area.\footnote{A non-trivial model is obtained by choosing $\vphi>0$. The stability is ensured by the usual isoperimetric inequality that implies that $A$ is bounded above at fixed $\ell$.}

More generally, from the gravitational point of view, the on-shell Eq.\ \eqref{eomonbd1} shows that the condition \eqref{bcIIIset} yields
\be\label{gravbcIII} n^{\mu}\partial_{\mu}\Phi = \la\, .\ee
This suggests that our b.c.\ III should be equivalent, in the second order formulation, to imposing the Neumann b.c.\ \eqref{gravbcIII} on $\Phi$ on top of fixing the integrated extrinsic curvature $\Upsilon = \oint k\d s$. As a simple consistency check, one can straightforwardly show that the equations of motion then implies that $\Phi=\la k=\text{constant}$ on the boundary and thus in particular that $\chi^{\mu}=-\la t^{\mu}$. This interpretation seems to be confirmed by the very recent analysis in \cite{Goel}; our b.c.\ III correspond to the b.c.\ called $\text{NN}^{*}$ in \cite{Goel}. 

\paragraph{Class IV}

The field $\Phi$ is fixed on the boundary. After integrating out $B$ we are left with the boundary action
\be\label{SFour2}  S_{\text{IV}} = \frac{\varphi}{8\pi}\oint_{\mathcal C}\bigl(\omega-\bar\omega\bigr)-\frac{1}{8\pi}\oint_{\mathcal C}\delta_{ab}\chi^{a} e^{b} = -\frac{\vphi}{8\pi}\oint_{\mathcal C}k\,\d s + \frac{\la}{8\pi}\,\ell\, .\ee
This is similar to b.c.\ II., i.e.\ JT gravity, but now $\ell$ is not fixed and we have a boundary cosmological constant term. Clearly, the partition functions of the models for b.c.\ II and b.c.\ IV are related by a Legendre transform. In a holographic set-up, the partition function with b.c.\ II on a disk is interpreted as being related to the partition function of a quantum mechanical system in the canonical ensemble at bare temperature $1/\ell$; by Legendre transforming we go to the microcanonical ensemble, with $\la$-dependent energy $-\la/(8\pi)$ (one may add an arbitrary zero-point energy to get the physical energy; this corresponds to a standard counterterm in JT gravity).

From the gravitational point of view, it is natural to assume that this b.c.\ amounts to fixing both $\Phi=\vphi$ and $n^{\mu}\partial_{\mu}\Phi=\la$ on the boundary. This seems to be confirmed by the analysis of the b.c.\ called DN in \cite{Goel}.

\subsection{\label{btopoSec}Interpretation and properties of the b.\ c.\ I}

The goal of this last subsection is to discuss the geometry associated with our b.c.\ I and to work out some consequences.

Let us recall some basic facts about the case of geodesic boundaries. The moduli space $\mathcal M_{g,b}$ of hyperbolic metrics is then of real dimension $6g-6+2b$. The existence of this space follows rather straightforwardly from the usual uniformization theorem for closed Riemann surfaces.\footnote{The idea is to use the so-called doubling construction. One builds a closed surface $\tilde{\mani}$, which is a double cover of $\mani$, by gluing two copies of $\mani$ along the boundaries, with an involution that acts as complex conjugation. One then uses the standard uniformization theorem for $\tilde{\mani}$.} The partition function coincides with the Weil-Petersson volume $\mathcal V_{g,b}(\ell_{1},\ldots,\ell_{b})$ of $\mathcal M_{g,b}$ considered by Mirzakhani. Remarkably, these volumes are polynomials in the lengths $\ell_{i}$ whose coefficients are famously related to the intersection numbers on the moduli space of punctured surfaces \cite{Mirzakhani}.

If we now consider boundaries with non-zero extrinsic curvature, it is not immediately obvious what the moduli space looks like or even if such metrics exist at all in general. Some aspects of the problem have been considered in the mathematical literature. References that we found particularly useful include \cite{Brendle,Rosenberg, Lopez,Rupflin} and references therein. A typical approach is to start from a metric $g_{0}$ with geodesic boundaries and to look for a metric $g=e^{2\sigma}g_{0}$ in the same conformal class having the required extrinsic curvature $k_{i}$ on the $i^{\text{th}}$ boundary component $\mathcal C_{i}$. This amounts to solving the differential equation
\be\label{diffeqforcstk} \Delta_{0}\sigma = 1-e^{2\sigma}\ \text{on}\ \mani\, ,\quad k_{i}e^{\sigma}=n_{0}^{\mu}\partial_{\mu}\sigma\ \text{on}\ \mathcal C_{i}\, .\ee
This approach seems to work well when $k_{i}<1$. However, it is not particularly illuminating for our purposes. We shall instead analyse the problem from the point of view of the $\PslR$ gauge theoretic formulation. Some results found in the mathematical literature can be easily recovered in this way, together with additional insights.

\subsubsection{\label{cutglueSec}Cutting and gluing cylinders}

\begin{figure}
\centerline{\includegraphics[width=6in]{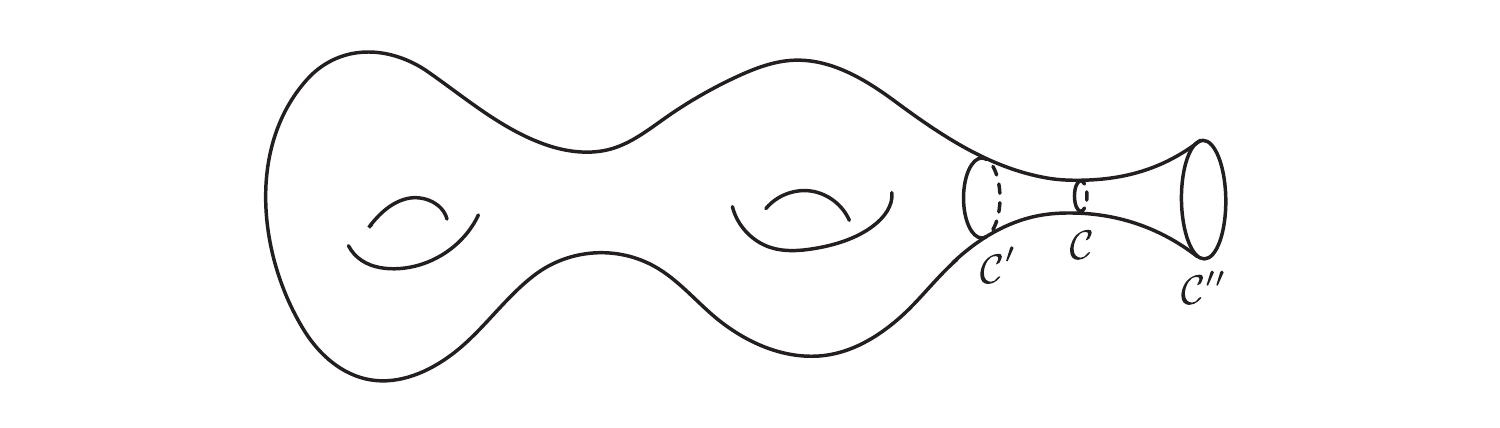}}
\caption{Example of a genus two surface $\mani$ with one boundary. The boundary is originally at $\mathcal C$, which has length $\ell$ and constant extrinsic curvature $k$, but it can be moved to $\mathcal C'$ or $\mathcal C''$, which have $(\ell', k')$ and $(\ell'',k'')$, without changing the partition function. As explained in the main text, $\ell^{2}(1-k^{2})$ must be invariant and the partition function depends only on this combination of parameters.}\label{excisecylfig}
\end{figure}

Near a circle boundary $\mathcal C$ characterized by $(\ell,k)$, the surface looks like a cylinder bounded on one side, say to the right, by $\mathcal C$. Any simple closed curve $\mathcal C'$ circling once around the cylinder is homotopic to the boundary.\footnote{It is convenient to work with free homotopy classes, i.e.\ homotopy with no base point, or equivalently with conjugacy classes in $\pi_{1}(\mani)$.} Assume that, for any metric having $(\ell,k)$ boundary condition on $\mathcal C$, we can find such a $\mathcal C'$ having constant extrinsic curvature $k'$ and length $\ell'$. By cutting the tubular region between $\mathcal C'$ and $\mathcal C$, we build a bijection between metrics having different boundary parameters $(\ell,k)$ and $(\ell',k')$. Moreover, since the on-shell bulk action is zero and there is no boundary action for b.c.\ I either, it is clear that the partition function $Z$ evaluated for the parameters $(\ell,k)$ must be equal to the partition function evaluated for the parameters $(\ell',k')$. A similar reasoning can be made by extending the tubular region to the right of $\mathcal C$. If a curve $\mathcal C''$ having constant extrinsic curvature $k''$ and length $\ell''$ can be found in the extended region, then we also conclude that $Z$ evaluated for $(\ell,k)$ is equal to $Z$ evaluated for $(\ell'',k'')$. This procedure is illustrated on Fig.\ \ref{excisecylfig}.

We now have the following simple lemma.

\noindent\emph{Lemma}: If two curves $\mathcal C$ and $\mathcal C'$ of lengths $\ell$ and $\ell'$ and constant extrinsic curvatures $k$ and $k'$ are homotopic to each other, then the  condition
\be\label{klkplprel} \ell^{2}(1-k^{2})=\ell'^{2}(1-k'^{2})\ee
must be satisfied. Moreover, if $k^{2}\geq 1$, then $k$ and $k'$ must have the same sign.

To prove the lemma, the idea is to compute the holonomies of the $\PslR$ connection $A$ along the contours $\mathcal C$ and $\mathcal C'$. To do that, we use local frame rotations to choose $e^{1}=t$ along $\mathcal C$. One then gets $k = -\omega(t)$, which is a special case of \eqref{Komega} in the gauge $\bar\omega=0$. The holonomy along $\mathcal C$ is thus 
\be\label{holoC} \text{Hol}(\mathcal C) =\text P \exp\Bigl( -\oint_{\mathcal C}A\Bigr) = \text P \exp\Bigl(\oint_{\mathcal C} \bigl(k J - K_{1}\bigr)\d s\Bigl)\, .\ee
Since we specialize to a case for which the extrinsic curvature is constant along the contour, we get
\be\label{Holocstk} \text{Hol}(\mathcal C) = e^{\ell(k J - K_{1})}\, ,\quad \text{Hol}(\mathcal C') = e^{\ell' (k' J - K_{1})}\, .\ee
But because the connection $A$ is flat and $\mathcal C$ and $\mathcal C'$ are homotopic, $\text{Hol}(\mathcal C)$ and $\text{Hol}(\mathcal C')$ must be conjugage to each other. Recall that for an arbitrary $\PslR$ element $\exp( w^{0} J - w^{1}K_{1} - w^{2} K_{2})$, the triplet $w=(w^{0},w^{1},w^{2})$ behaves like a Lorentz three-vector under conjugation. The conjugacy classes are thus classified by the three-norm $w^{2}=-(w^{0})^{2}+(w^{1})^{2}+(w^{2})^{2}$ and by the sign of $w^{0}$ when $w^{2}\leq 0$.\footnote{Time-like, light-like and space-like vectors $w$ are associated with so-called elliptic, parabolic and hyperbolic elements of $\PslR$.} The lemma then follows from this classification.

The above discussion shows that we have five qualitatively distinct cases to consider, $k<-1$, $k=-1$, $-1<k<1$, $k=1$ and $k>1$, associated with the conjugacy classes of $\text{Hol}(\mathcal C)$ in $\PslR$. It also suggests that, for each case, the partition function depends only on the combination of parameters $\ell^{2}(1-k^{2})$. We are now going to prove this statement and provide simple geometric interpretations of the various possibilities, by analysing the five cases one by one.  

\subsubsection{The case $-1<k<1$} 

\paragraph{Using the collar lemma} Let us first assume that we start from a geodesic boundary $\mathcal C$ of length $\bar \ell$. A standard result \cite{Randol}, known as the collar lemma, shows that the vicinity of $\mathcal C$ is topologically a cylinder with a metric
\be\label{methyperbolicbd} \d s^{2} = \frac{\d\tau^{2}+\d\sigma^{2}}{\cos^{2}\sigma}\,\cvp\quad -\sigma_{-}<\sigma\leq 0\, ,\ \tau\equiv\tau + \bar\ell\, .\ee
The cylinder extends at least all the way to $\sigma=\sigma_{-}$, with
\be\label{sigmaminus} \sigma_{-} = \frac{\pi}{2}-\arctan\sinh\frac{\bar\ell}{2}\, \cvp\ee
without intersecting neighbouring cylinders attached to other boundaries. Note that this cylinder is naturally described in terms of the usual strip representation of hyperbolic space, for which we periodically identify $\tau$ and $\tau+\bar \ell$. Closed curves $\sigma=\text{constant}$ have
\be\label{klhyper} \ell(\sigma) = \frac{\bar\ell}{\cos\sigma}\,\cvp\quad k(\sigma) = \sin\sigma\ee
with
\be\label{relOK1} \ell(\sigma)^{2}\bigl(1-k(\sigma)^{2}\bigr) = \bar\ell^{2}\ee
independent of $\sigma$. This is consistent with the discussion of the previous subsection, from which we also know that the partition function will not depend on the actual location of the boundary in the interval $-\sigma_{-}<\sigma<\pi/2$. In particular, all the choices of $(\ell, k)$ with $\ell^{2}(1-k^{2})=\bar\ell^{2}$ and for which $k(\sigma_{-}) = -1/\cosh(\bar\ell/2)<k<1$ yield the same partition function. 

The lower bound on the accessible values of $k$ is due to the lower bound in $\sigma$ provided by the collar lemma. This restriction can be easily waived. Indeed, if we start from a boundary $(\ell,k)$, for any $k$ such that $|k|<1$, then its immediate vicinity will also be a cylinder with metric \eqref{methyperbolicbd}, the boundary being now at $\sigma_{\text b} = \arcsin k$ and the cylinder extending at least down to $\sigma = \sigma_{\text b}-\epsilon$, for some $\epsilon>0$. There is no obstacle to extending the cylinder to larger values of $\sigma$, without changing the partition function. In particular, any choice $-1<k<0$ boils down in this way to the geodesic case.

\paragraph{Using $\PslR$ as a metric-generating ``spontaneously broken'' symmetry}

As emphasized many times before, the $\PslR$ gauge symmetry is broken down to $\PslR_{\partial}$. In the case of b.c.\ of class I, this breaking is entirely due to the boundary conditions themselves, because there is no symmetry-breaking boundary term in the action. The symmetry breaking is thus very similar to a spontaneous breaking. In particular, if we start from a metric having boundary parameters $(\ell,k)$ on a particular boundary component $\mathcal C$, we can generate new metrics with new boundary parameters $(\ell',k')$ on $\mathcal C$ by acting with $\PslR$, \emph{without changing the partition function.} We are going to illustrate this idea below.

Let us pick a metric for which $\mathcal C$ has parameter $(\ell,k)$, $|k|<1$. Setting $\sigma_{\text b}=\arcsin k$, we can use the explicit form \eqref{methyperbolicbd}, at least in a neighbourhood $\mathcal C_{\epsilon}$, $\sigma_{\text b}-\epsilon<\sigma\leq\sigma_{\text b}$, of the boundary at $\sigma=\sigma_{\text b}$. This metric may be extended for $\sigma\in (\sigma_{\text b}-\epsilon,\pi/2)$. The associated flat $\PslR$ connection can be straightforwardly computed and reads
\be\label{Ahypercase} A =\omega J + e^{a}K_{a}= -J\tan\sigma\,\d\tau + \frac{1}{\cos\sigma}\bigl( K_{1}\d\sigma + K_{2}\d\tau\bigr)\, .\ee
Let us now consider a $\PslR=\text{SO}(2,1)_{+}$ gauge transformation, which is the identity in the bulk of $\mani$, except in $\mathcal C_{\epsilon}$ where it is a Lorentz boost in the $e^{2}$ direction, with a monotonic smooth speed profile $V(\sigma)$ such that
\be\label{Boostprofile}  V(\sigma) = \left\{
\begin{aligned} & 0 \quad\text{for}\ \sigma\leq \sigma_{\text b}-\epsilon\\
& v\quad\text{for}\ \sigma>\sigma_{\text b}-\epsilon'>\sigma_{\text b}-\epsilon\, .
\end{aligned}\right.\ee
This acts as
\be\label{Boostact1} \omega ' = \gamma(V)\bigl(\omega - V e^{2}\bigr)\, ,\quad e'^{1} = e^{1}+\gamma(V)^{2}\d V\, ,\quad e'^{2}=\gamma(V)\bigl( e^{2}-V\omega\bigr)\ee
where
\be\label{gammadefrel} \gamma(V) = \frac{1}{\sqrt{1-V^{2}}}\ee
is the usual relativistic factor.\footnote{Such a gauge transformation is an automorphism of the vector bundle $\mathscr E\rightarrow\mani$ since it does not change the transition functions. It corresponds to a global section of the associated $\PslR$ adjoint bundle.} This produces a new hyperbolic metric on $\mani$ which coincides with the original metric everywhere except on $\mathcal C_{\epsilon}$, where it reads
\be\label{merhyperafter} \d s'^{2} = \frac{1}{\cos^{2}\sigma}\Bigl[ \bigl(1+\gamma(V)^{2}\frac{\d V}{\d\sigma}\cos\sigma\bigr)^{2}\d\sigma^{2} + \gamma(V)^{2}\bigl(1+ V \sin\sigma\bigr)^{2}\d\tau^{2}\Bigr]\, . \ee
This metric may be extended to all values of $\sigma\in (\sigma_{\text b}-\epsilon,\pi/2)$ by setting $V=v$ for all $\sigma\geq\sigma_{\text b}$. The extrinsic curvature $k'(\sigma)$ and length $\ell'(\sigma)$ of the $\sigma=\text{constant}>\sigma_{\text b}-\epsilon'$ curves in the new metric are expressed in terms of the corresponding quantities for the old metric, Eq.\ \eqref{klhyper}, as
\be\label{boostkl1} k'(\sigma) = \frac{k(\sigma) + v}{1+v k(\sigma)}\,\cvp\quad \ell'(\sigma) = \gamma(v)\bigl(1+vk(\sigma)\bigr)\ell(\sigma)\, .\ee
The transformation law for $k$ is the same as the composition law for velocities in special relativity. One can check that
\be\label{klrelsig} \ell'(\sigma)^{2}\bigl(1-k'(\sigma)^{2}\bigr) = \ell(\sigma)^{2}\bigl(1-k(\sigma)^{2}\bigr)\, ,\ee
as expected. 

Clearly, starting from any $k\in (-1,1)$, we can get any other $k'\in (k,1)$ by using a boost with $v>0$. The metric \eqref{klrelsig} is then automatically smooth, because $\d V/\d\sigma >0$. By working a little bit harder, one can also show that, starting from any $k\in (0,1)$, we can get a geodesic, $k'=0$. We first use the generalized collar lemma, derived in \cite{Rupflin}, showing that the cylindrical metric can always be extended to an interval $-\epsilon\leq\sigma\leq\sigma_{\text b}=\arcsin k$ in this case. We thus need to construct a smooth profile $V(\sigma)$ satisfying 
\be\label{Boostprofile2}  V(\sigma) = \left\{
\begin{aligned} & 0 \quad\text{for}\ \sigma\leq -\epsilon\\
& v=-k\quad\text{for}\ \sigma>\sigma_{\text b}-\epsilon'\, ,
\end{aligned}\right.\ee
where $\epsilon'$ may be as small as we need, with the constraint $1+\gamma(V)^{2}(\d V/\d\sigma)\cos\sigma>0$ ensuring the smoothness of the new metric. One can obtain $V$ by slightly deforming $\tilde V(\sigma) = -\sin\sigma$, which satisfies $1+\gamma(\tilde V)^{2}(\d \tilde V/\d\sigma)\cos\sigma=0$, $\tilde V(0) = 0$ and $\tilde V(\sigma_{\text b}) = -k$.

The above discussion proves that the partition function does not depend on the parameters $\ell$ and $k$ independently, but only on the invariant combination
\be\label{lbardef} \bar\ell = \ell\sqrt{1-k^{2}}\,.\ee
Note that this result relies on the fact that the partition function does not change when we act with a $\PslR$ gauge transformation. A similar reasoning does not work for the other classes of b.c., due to the explicit symmetry-breaking boundary terms one must then include in the action.

\vfill\eject

\noindent\emph{Remarks}:

i) The above simple construction bypasses the non-trivial analysis of the system of equations \eqref{diffeqforcstk} to prove the existence of metrics with $k\not = 0$, starting from the well-known geodesic case.

ii) Of course, the boost $\PslR$ transformations that we have been using above are not equivalent to diffeomorphisms. In particular, they change the length and the extrinsic curvature of the boundary. They are $\PslR$ gauge transformations that do not belong to $\PslR_{\partial}$. 

iii) The metric \eqref{merhyperafter} in the region where $V=v=\text{constant}$ can be put into the canonical form \eqref{methyperbolicbd}, $\d s'^{2}=(\d\tilde\sigma^{2}+\d\tau^{2})/\cos^{2}\tilde\sigma$, by the change of coordinate
\be\label{sigtosigtilde1} \sin\tilde\sigma = \frac{\sin\sigma + v}{1+v\sin\sigma}\, \cdotp\ee
This is a diffeomorphism $(-\pi/2,\pi/2)\rightarrow (-\pi/2,\pi/2)$ but it is \emph{not} a diffeomorphism of $\mani$ because it moves the boundary. We find again the crucial distinction between $\PslR$ gauge transformation and diffeomorphisms.

\paragraph{The funnel geometry}

The cylinder with metric \eqref{methyperbolicbd} is called a \emph{funnel}. This metric depends on a unique parameter $\bar\ell$, corresponding to the length of the geodesics it contains. The above discussion implies that imposing boundary parameters $(\ell,k)$ with $|k|<1$ on a boundary component $\mathcal C$ of $\mani$ is equivalent to attaching a funnel with $\bar\ell=\ell\sqrt{1-k^{2}}$ to $\mani$. Without loss of generality, one may assume that the boundary is the geodesic of length $\bar\ell$, but the funnel may equivalently be cut at any value of $\sigma\in (-\pi/2+\arctan\sinh(\bar\ell/2),\pi/2)$. In particular, we may extend the surface ``to infinity,'' $\sigma\rightarrow\pi/2$, taking the limit $\ell\rightarrow\infty$, $k\rightarrow 1^{-}$ at fixed $\bar\ell$.

\paragraph{The $\bar\ell\rightarrow 0$ limit}

Let us now study the limit $\bar\ell\rightarrow 0$ of the invariant parameter \eqref{lbardef}. From above, we know that we can work with a geodesic boundary of length $\bar\ell$, without loss of generality. The metric near the boundary is \eqref{methyperbolicbd}. Intuitively, when $\bar\ell\rightarrow 0$, we are ``pinching'' the surface. But what is really the nature of this pinch? 

To understand what is happening, it is convenient to introduce new radial-like and angular coordinates $\hat\rho$ and $\hat\theta$ defined by
\be\label{rhohat} \sigma = -\frac{\pi}{2} -\frac{\bar\ell}{2\pi}\ln\hat\rho\, ,\quad \hat\theta = \frac{2\pi\tau}{\bar\ell}\,\cdotp\ee
The angular variable $\hat\theta$ is $2\pi$ periodic and 
\be\label{varinthatrho} e^{-\pi^{2}/\bar\ell}\leq\hat\rho<\hat\rho_{-}=e^{-\pi^{2}/\bar\ell+2\pi\sigma_{-}/\bar\ell}<1\, .\ee
In these new variables, the metric \eqref{methyperbolicbd} reads
\be\label{newmethyperbolicbd} \d s^{2} = \Bigl(\frac{\bar\ell}{2\pi}\Bigr)^{2}
\frac{\d\hat\rho^{2} + \hat\rho^{2}\d\hat\theta^{2}}{\hat\rho^{2}\sin^{2}[\frac{\bar\ell}{2\pi}\ln\hat\rho]}\,\cdotp\ee
This metric is valid at least up to $\hat\rho = \hat\rho_{-}$ where the cylinder attaches to the rest of the surface. The crucial input in now the explicit expression \eqref{sigmaminus} for $\sigma_{-}$, or equivalently for $\hat\rho_{-}$, given by the collar lemma. This expression shows that $\lim_{\bar\ell\rightarrow 0}\hat\rho_{-}\rightarrow e^{-\pi}$ is finite. It thus makes sense to take the $\bar\ell\rightarrow 0$ limit of \eqref{newmethyperbolicbd} at fixed $\hat\rho$, yielding
\be\label{cuspmetaslimit} \d s^{2} = \frac{\d\hat\rho^{2} + \hat\rho^{2}\d\hat\theta^{2}}{\hat\rho^{2}\ln^{2}\hat\rho}\,\cdotp\ee
This is the metric for a so-called \emph{cusp} singularity. Note that the geodesic boundary that was originally at $\hat\rho = e^{-\pi^{2}/\bar\ell}$ has been pushed at infinite distance from the rest of the surface. In fact, after the limit is taken, the geodesic sits beyond the cusp and thus disappears.

\paragraph{R\'esum\'e} When $-1<k<1$, the partition function depends only on the invariant parameter $\bar\ell = \ell\sqrt{1-k^{2}}$. Without loss of generality, one may assume that the boundary is a geodesic of length $\bar\ell$. More generally, the geometry near the boundary is a funnel containing a geodesic of length $\bar\ell$. The limit $\bar\ell\rightarrow 0$ corresponds to the insertion of a cusp singularity.

\subsubsection{The case $k<-1$}

We can repeat the logic used for the case $k\in (-1,1)$. The vicinity of a $k<-1$ boundary is a cylindrical region with a metric
\be\label{metellipticb}\d s^{2} = \frac{4}{(1-r^{2})^{2}}\bigl(\d r^{2} + r^{2}\d\phi^{2}\bigr)\, ,\quad r_{\text b}\leq r < r_{\text b}+\epsilon\, ,\ \phi\equiv\phi + 2\pi\alpha\, ,\ee
the boundary being at $r=r_{\text b}$ and for some $\epsilon>0$. This cylinder is naturally described in terms of the Poincar\'e disk representation of hyperbolic space, but for which the angular variable $\phi$ is not necessarily $2\pi$ periodic. Closed curves $r=\text{constant}$ have
\be\label{klellip} \ell(r) = \frac{4\pi r}{1-r^{2}}\,\alpha\, ,\quad k(r) = -\frac{1+r^{2}}{2r}\, \cdotp\ee
The sign of $k(r)$ corresponds to the case where the curve is seen as bounding a surface that extends outward, consistently with the condition $k<-1$. One can check that
\be\label{relOK2} \ell(r)^{2}\bigl(k(r)^{2}-1\bigr) = (2\pi\alpha)^{2}\ee
is independent of $r$, as expected. 

The $\PslR$ connection near the boundary now reads
\be\label{Aellipcase} A = \omega J + e^{a}K_{a} = -J\,\frac{1+r^{2}}{1-r^{2}}\,\d\phi + \frac{2}{1-r^{2}}\bigl( K_{1}\,\d r +K_{2}\,r\d\phi\bigr)\, .\ee
We can act with the boost $\PslR$ transformation \eqref{Boostact1}, for a monotonic velocity profile $V$ satisfying 
\be\label{Boostprofile3}  V(r) = \left\{
\begin{aligned} & 0 \quad\text{for}\ r\geq r_{\text b}+\epsilon\\
& v\quad\text{for}\ r_{\text b}\leq r<r_{\text b}+\epsilon'<r_{\text b}+\epsilon\, ,
\end{aligned}\right.\ee
to generate a new metric
\be\label{metellipafter} \d s'^{2} = \frac{4}{(1-r^{2})^{2}}\biggl[
\Bigl(1+\frac{1}{2}(1-r^{2})\gamma(V)^{2}\frac{\d V}{\d r}\Bigr)^{2} \d r^{2} 
 + \gamma(V)^{2}\bigl(r+\frac{1}{2}(1+r^{2})V\bigr)^{2}\d\phi^{2}\biggr]\, .\ee
The extrinsic curvature $k'(r)$ and length $\ell'(r)$ of the $r=\text{constant}$ curves in the new metric, in the region $r_{\text b}\leq r<r_{\text b}+\epsilon'$, are expressed in terms of the corresponding quantities for the old metric, Eq.\ \eqref{klellip}, as
\be\label{boostkl2} k'(r) = \frac{k(r) - v}{1-v k(r)}\,\cvp\quad \ell'(r) = \gamma(v)\bigl(1-vk(r)\bigr)\ell(r)\, ,\ee
similarly to the previous case \eqref{boostkl1}. Of course,
\be\label{klrelr} \ell'(r)^{2}\bigl(k'(r)^{2}-1\bigr) = \ell(r)^{2}\bigl(k(r)^{2}-1\bigr)\, .\ee
Clearly, starting from any $k<-1$, we can get any other $k'<k$ by using a boost with $v\in(1/k,0)$. This yields a smooth metric \eqref{metellipafter}, because $\d V/\d r >0$ and it can also be checked straightforwardly that $r+\frac{1}{2}(1+r^{2})V(r) = r(1- k(r)V(r))>0$ for all $r\in [ r_{\text b},r_{\text b}+\epsilon]$.\footnote{On the other hand, we cannot find a profile that yields a smooth metric if $v<1/k$; this would have produced a metric with $k'>1$!}

From this we deduce that the partition function does not depend on $\ell$ and $k$ independently, but only on the invariant combination 
\be\label{alphadef} 2\pi\alpha = \ell\sqrt{k^{2}-1}\, .\ee
In particular, without loss of generality, we can take the limit $r_{\text b}\rightarrow 0$, or equivalently $\ell\rightarrow 0$, $k\rightarrow -\infty$ at $\alpha$ fixed. \emph{This amounts to inserting a conical singularity of angle $2\pi\alpha$, i.e.\ of deficit angle $2\pi(1-\alpha)$, on $\mani$}. 

Setting $\hat\rho = r^{1/\alpha}$, $\hat\theta = \phi/\alpha$, the metric is brought into the form
\be\label{conemetricstandard} \d s^{2} = \frac{4\alpha^{2}\hat\rho^{2\alpha -2}}{(1-\rho^{2\alpha})^{2}}\bigl(\d\hat\rho^{2} + \hat\rho^{2}\d\hat\theta^{2}\bigr)\, ,\quad \hat\theta\equiv\hat\theta + 2\pi\, .\ee
This is particularly convenient to study the limit $\alpha\rightarrow 0$: we get a cusp singularity \eqref{cuspmetaslimit}.

\paragraph{R\'esum\'e} When $k<-1$, the partition function depends only on the invariant parameter $2\pi\alpha = \ell\sqrt{k^{2}-1}$. Without loss of generality, one may assume that a conical singularity of angle $2\pi\alpha$ has been inserted on $\mani$. The limit $\alpha\rightarrow 0$ corresponds to the insertion of a cusp singularity.

\subsubsection{The case $k=-1$}

The vicinity of a $k=-1$ boundary is a cylindrical region with a metric
\be\label{metparabolic} \d s^{2} = \frac{\d u^{2}+\d v^{2}}{v^{2}}\, \cvp\quad v_{\text b}-\epsilon< v \leq v_{\text b}\, ,\ u\equiv u + \ell\, ,\ee
the boundary, of length $\ell(v_{\text b})=\ell/v_{\text b}$, being at $v=v_{\text b}$. It is straightforward to check that we can use a boost $\PslR$ transformation in the direction 1 (similar to \eqref{Boostact1}, but with $e^{2}$ and $e'^{2}$ replaced by $e^{1}$ and $e'^{1}$) to reduce the boundary length to an arbitrarily small value; in particular, we can take the limit $\ell(v_{\text b})\rightarrow 0$. Setting $u=\frac{\ell}{2\pi}\hat\theta$ and $v=-\frac{\ell}{2\pi}\ln\hat\rho$, one finds that this is equivalent to creating a cusp singularity. This result is not surprising. The case $k=-1$ can be seen as the limit $k\rightarrow -1^{+}$ at fixed $\ell$ of the case $k\in (-1,1)$, or equivalently as the limit $k\rightarrow -1^{-}$ at fixed $\ell$ of the case $k<-1$, which both yield a cusp.

\paragraph{R\'esum\'e} Having a boundary with $k=-1$ is equivalent to inserting a cusp singularity on $\mani$. 

\subsubsection{\label{kgreatSec}The case $k>1$}

We can mimic the analysis made for $k<-1$ and deduce that the partition function depends only on the invariant parameter \eqref{alphadef}. However, there is a crucial difference. The metric in the cylindrical neighborhood of the boundary takes the form \eqref{metellipticb}, but the surface now lies in the interior, for $r\leq r_{\text b}$. It is thus impossible to use the $r_{\text b}\rightarrow 0$ limit. But we may take the opposite limit of large $r_{\text b}$ to create a large asymptotic region
\be\label{metellpos} \d s^{2} = \frac{4}{(1-r^{2})^{2}}\bigl(\d r^{2} + r^{2}\d\phi^{2}\bigr)\, ,\quad r_{\text b}-\epsilon< r \leq r_{\text b}\, ,\ \phi\equiv\phi + 2\pi\alpha\, .\ee

A very useful recent reference to understand the metrics for which certain boundaries have $k>1$ is \cite{Lopez}. The basic idea, first put forward by Rosenberg in \cite{Rosenberg} (see also \cite{Sun}) is the following.

\noindent\emph{Lemma} Let $\mathcal C_{1}$ and $\mathcal C_{2}$ be two boundary components, of fixed extrinsic curvatures $k_{1}$ and $k_{2}$. Assume that there exists a geodesic starting on $\mathcal C_{1}$ and ending on $\mathcal C_{2}$ that minimizes, at least locally, the distance between $\mathcal C_{1}$ and $\mathcal C_{2}$. Then $k_{1}>1$ implies that $k_{2}<-k_{1}$.

\begin{figure}
\centerline{\includegraphics[width=6in]{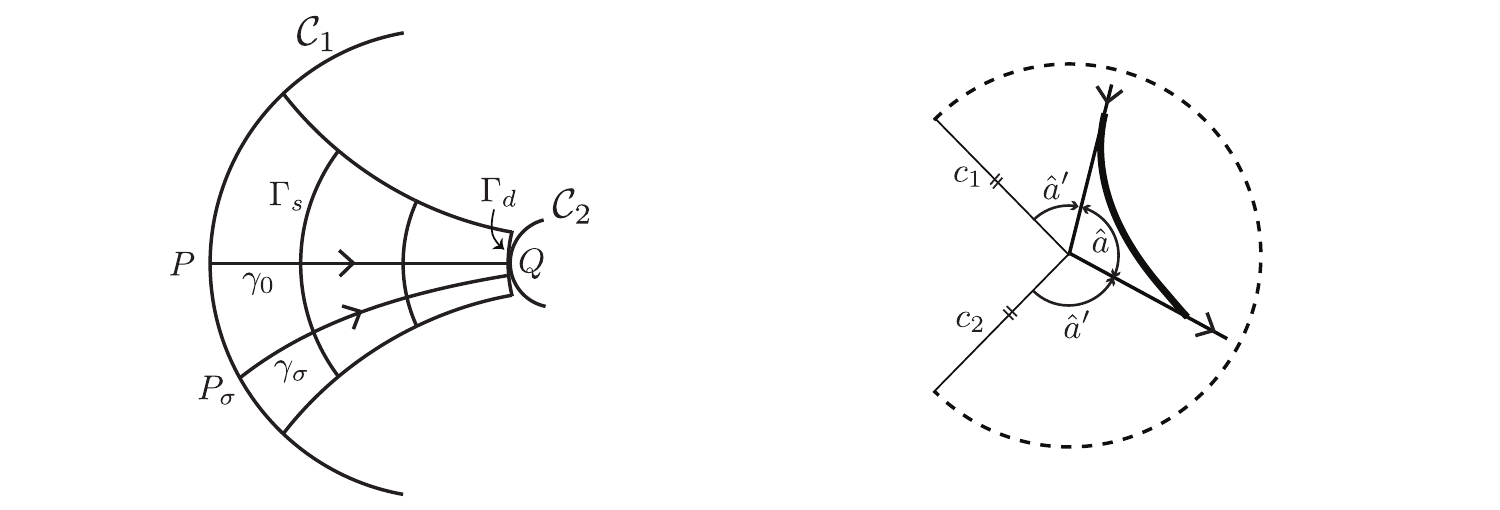}}
\caption{Left inset: boundary components $\mathcal C_{1}$ and $\mathcal C_{2}$ of extrinsic curvatures $k_{1}$ and $k_{2}$ joined by a distance-minimizing geodesic $\gamma_{0}$. If $k_{1}>1$, the curves $\Gamma_{s}$, defined in the main text, have increasing extrinsic curvatures. The position of $\Gamma_{d}$ with respect to $\mathcal C_{2}$ then implies that $k_{2}<-k_{1}$. Right inset: conical singularities of angles less than $2\pi$ never attract distance minimizing geodesics like $\gamma_{0}$ on the left inset. If they were, their length could be reduced by smoothing the corner as indicated (thick curve). This always works because the geodesics $c_{1}$ and $c_{2}$ (thin lines) are identified to form the cone and one of the angles $\hat a$ or $\hat a'$ is then always less than $\pi$.}\label{lemmaRosenbergfig}
\end{figure}

The proof is based on the following idea, which is illustrated on Fig.\ \ref{lemmaRosenbergfig}. One considers a family of geodesics $\gamma_{\sigma}(s)$, $s$ being the arc length along $\gamma_{\sigma}$. The central geodesic $\gamma_{0}$ minimizes the distance between $\mathcal C_{1}$ and $\mathcal C_{2}$, with $\gamma_{0}(0)=P\in\mathcal C_{1}$ and $\gamma_{0}(d)=Q\in\mathcal C_{2}$. The $\gamma_{\sigma}$ are nearby geodesics that starts orthogonally to $\mathcal C_{1}$ at $\gamma_{\sigma}(0)=P_{\sigma}\in\mathcal C_{1}$, where $\sigma$ is the distance from $P$ to $P_{\sigma}$ along $\mathcal C_{1}$. The parameter $\sigma$ varies in an arbitrarily small interval $(-\epsilon,\epsilon)$. The curve $\Gamma_{0}:\sigma\mapsto\gamma_{\sigma}(0)$ is thus a small portion of $\mathcal C_{1}$ for which $\sigma$ is the length parameter. Note that since $\gamma_{0}$ minimizes the distance, there is no conjugate point to $\gamma_{0}(0)$ along $\gamma_{0}$. One can then show rather straightforwardly, using the Jacobi equation, that the extrinsic curvature $k(s)$ of the curve $\Gamma_{s}:\sigma\mapsto\gamma_{\sigma}(s)$ satisfies the equation
\be\label{Jacobiext} \frac{\d k}{\d s} = k^{2} + \frac{R}{2} = k^{2}-1\, .\ee
It is thus a strictly increasing function of $s$ as soon as $k(0)=k_{1}>1$. Now, since $\gamma_{0}(d)=Q\in\mathcal C_{2}$ and $\gamma_{0}$ minimizes the distance between $\mathcal C_{1}$ and $\mathcal C_{2}$, the geodesics $\gamma_{\sigma}(s)$ for $\sigma\not = 0$ can be extended to $s=d$ without intersecting $\mathcal C_{2}$. The curve $\Gamma_{d}:\sigma\mapsto\gamma_{\sigma}(d)$ is thus well-defined (it belongs to the surface) and is tangent to $\mathcal C_{2}$ at $\gamma_{0}(d)$. But the position of $\Gamma_{d}$ with respect to $\mathcal C_{2}$ implies that $k(d)<-k_{2}$, the minus sign taking into account the orientation of the normal vector used to compute $k_{2}$. Since $k(d)>k_{1}$, we conclude.

The lemma has rather strong consequences on the possible geometries. For instance, it implies immediately that if $k_{i}>1$ on all the boundaries, then there is actually only one boundary and $\mani=\disk$ is the disk (the proof goes as follows; one first considers a minimizing geodesic between a given circle component and all the others to deduce that there can be only one boundary; one then considers a minimizing geodesic that starts and ends on the unique boundary component by going through a handle of the surface to deduce that there cannot be handles).

This argument can be generalized along the following lines. Consider first a surface containing both $k>1$ boundaries and $k\in [-1,1)$ boundaries. As we have seen before, the $k\in [-1,1)$ boundaries can be pushed to very large distances, by extending the funnels they belong to or by creating cusps. So these boundaries do not interfere with the minimizing geodesics in the argument above and we can repeat it. We deduce that the only surfaces of this kind have a unique $k>1$ boundary and genus zero. Consider now a surface containing conical singularities as well. These are at finite distance; they may attract the path minimizing the distance between boundaries and ruin the validity of the above reasonings. However, this will never happen if the angles of the conical singularities are constrained to be less than $2\pi$.\footnote{Note that this is not the same condition that was used in \cite{Wittensummer}. In \cite{Wittensummer} the existence of a pair of pants decomposition of the surface was required. This yields the stronger constraint that the angles of the cones must be less than $\pi$.} This is illustrated on the right inset of Fig.\ \ref{lemmaRosenbergfig}. If a length minimizing curve passes through the conical point of angle less than $2\pi$, one of the angles ($\hat a$ or $\hat a'$ on the figure) between the incoming and outgoing paths must be less than $\pi$; the curve can then always be shortened by smoothing the corner, which is a contradiction. We thus obtain the following result: \emph{surfaces with $k>1$ boundaries must necessarily be planar and contain a unique $k>1$ boundary component, except if they also contain cones of angles strictly greater than $2\pi$.}

\begin{figure}
\centerline{\includegraphics[width=6in]{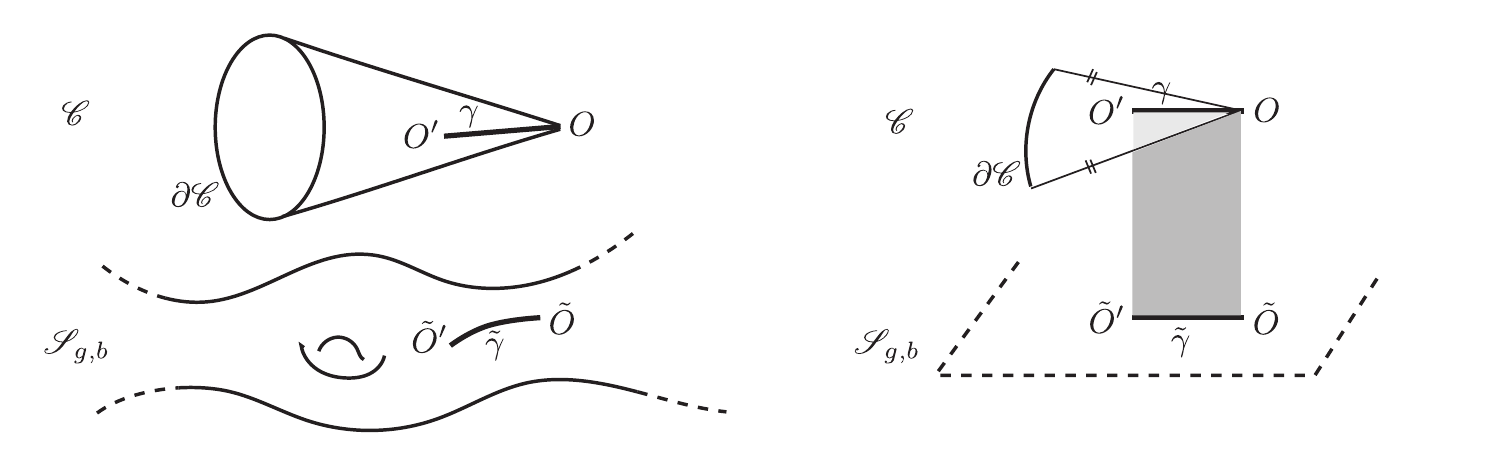}}
\caption{Construction of surfaces with an arbitrary number of boundary components having $k>1$. By gluing together a cone $\mathscr C$ (that may also be a disk with no conical singularity at the origin) and an arbitrary surface $\mathscr S_{g,b}$ along a short geodesic (thick lines $OO'$ and $\tilde O\tilde O'$), we get a new surface with one more $k>1$ boundary and at least one conical singularity of angle strictly greater than $2\pi$. To facilitate the drawing, the points $\tilde O$ and $\tilde O'$ are chosen to be smooth. Left inset: surface sketch. Right inset: local two-sheeted representation; the two sheets are glued through the shaded area.}\label{FigDima}
\end{figure}

On the other hand, if we allow cones with angles strictly greater than $2\pi$, surfaces with an arbitrary number of $k>1$ circle boundaries and of arbitrary genera can be constructed. One way to proceed, which is illustrated on Fig.\ \ref{FigDima}, is as follows.\footnote{We are grateful to Dima Panov for suggesting a construction along these lines.} We start from a cone $\mathscr C$ with a conical singularity of angle $2\pi\alpha$ at its center $O$ (the special case of the disk, $\alpha=1$, is perfectly allowed). This corresponds to the geometry \eqref{metellpos}, with a boundary at $r=r_{\text b}$, continued to $r=0$. We pick another point $O'$ on the cone, say at $r=r_{0}<r_{\text b}$, $\theta=0$ and we call $\gamma$ the geodesic of length $L =2\argtanh r_{0}$ joining $O$ and $O'$. We also consider a surface $\mathscr S_{g,b}$, of genus $g$, with $b$ boundaries, that may also contain conical singularities. We pick two points $\tilde O$ and $\tilde O'$ on this surface, joined by a geodesic $\tilde\gamma$ of length $L$. This can always be done, as long as $L$ is small enough. We might have conical singularities of angles $2\pi\tilde\alpha$ and $2\pi\tilde\alpha'$ at $\tilde O$ and $\tilde O'$ respectively. We then construct a new surface $\mathscr S_{g,b+1}$ by gluing $\mathscr C$ and $\mathscr S_{g,b}$ along $\gamma=\tilde\gamma$. The new surface has the same genus as $\mathscr S_{g,b}$, two conical singularities at $O=\tilde O$ and $O'=\tilde O'$ of angles $2\pi(\alpha+\tilde\alpha)$ and $2\pi(1+\tilde\alpha')$ respectively and, most importantly, one new $k>1$ boundary coming from the boundary of $\disk$. There is no contradiction with our previous claims because the angle of the cone at $O'=\tilde O'$ is necessarily greater than $2\pi$.\footnote{One might be tempted to generalize the construction by putting a conical singularity of angle $2\pi\alpha'<2\pi$ at the point $O'$, in order to reduce the final cone angle from $2\pi (1+\tilde\alpha')$ to $2\pi (\alpha'+\tilde\alpha')$, which may be less than $2\pi$. But putting a cone at $O'$ is not allowed because it  creates a cusp on the boundary.} By adjusting $\alpha+\tilde\alpha=1$, we may make the point $O=\tilde O$ smooth if we wish. By repeating this procedure, one can add as many $k>1$ boundaries as one wishes to the surface.

Considering $k>1$ boundaries thus yield an interesting generalization of the cases already studied in the literature. The simplest possible set-up is when we apply boundary conditions of class I on all boundary components. The partition function
\be\label{Zgenp} Z = Z(\ell_{1},\ell_{2},\ldots; \alpha_{1},\ldots;\tilde\alpha_{1},\ldots)\ee
depends on the lengths $\ell_{1},\ell_{2}$, etc., of the geodesic boundaries, on the angles $2\pi\alpha_{1},2\pi\alpha_{2}$, etc., of the conical singularities and on the invariant parameters $2\pi\tilde\alpha_{1} = \tilde\ell_{1}\sqrt{k_{1}^{2}-1}$, etc., associated with the $k>1$ boundaries. These quantitities are non-trivial generalizations of the volumes considered by Mirzakhani. In particular, because cones of angles greater than $2\pi$ (and thus of $\pi$) are necessarily present, the associated surfaces do not have a pair of pants decomposition, as recently emphasized in \cite{Wittensummer}. The computation of $Z$ in these cases is an interesting and challenging problem for the future.

\subsubsection{The case $k=1$}

We shall be brief here, in particular because this case can be seen as the limit $k\rightarrow 1^{+}$ at fixed $\ell$ of the case $k>1$.

One can derive a new version of the lemma used in the previous subsection, by using the same argument, in particular eq.\ \eqref{Jacobiext}:

\noindent\emph{Lemma} Let $\mathcal C_{1}$ and $\mathcal C_{2}$ be two boundary components, of fixed extrinsic curvatures $k_{1}=1$ and $k_{2}$. Assume that there exists a geodesic starting on $\mathcal C_{1}$ and ending on $\mathcal C_{2}$ that minimizes, at least locally, the distance between $\mathcal C_{1}$ and $\mathcal C_{2}$. Then $k_{2}=-1$.

This implies that the only geometry with $k=1$ boundaries is a cusp, Eq.\ \eqref{cuspmetaslimit}, extending up to a unique finite length boundary; the exception is when we allow for conical singularities of angles strictly greater than $2\pi$. In this case, we can use a gluing procedure analogous to the one depicted in Fig.\ \ref{FigDima} to build surfaces with complicated topologies and an arbitrary number of $k=1$ boundaries. The cone $\mathscr C$ is replaced by the cusp. The geodesic $\gamma$ is either of finite length and can then be glued to a geodesic $\tilde \gamma$ of the same length on an arbitrary surface $\mathscr S_{g,b}$; this adds two conical singularities of angles greater than $2\pi$, a cusp and a $k=1$ boundary to $\mathscr S_{g,b}$. Or it goes all the way down to the cusp, which may then be merged to another cusp on $\mathscr S_{g,b}$; this adds a conical singularity of angle greater than $2\pi$ and a $k=1$ boundary to $\mathscr S_{g,b}$.

%

%
\subsection*{Acknowledgments}

I am grateful to Joel Fine for an illuminating exchange on the case $k>1$ and to Dima Panov for suggesting the construction depicted on Fig.\ \ref{FigDima}. Section \ref{kgreatSec} would not have been the same without them. 

This work is supported in part by the Belgian Fonds National de la Recherche Scientifique FNRS (convention IISN 4.4503.15).



\begin{thebibliography}{99}

\bibitem{Polyakov}{A.M.~Polyakov, ``Quantum Geometry of Bosonic Strings,'' \plb{103}{1981}{207}.}
    
\bibitem{JTpapers}{C.~Teitelboim, ``Gravitation and Hamiltonian Structure in Two Space-Time Dimensions,'' \plb{126}{1983}{41},\\
R.~Jackiw, ``Lower-Dimensional Gravity,'' \npb{252}{1985}{343}.}

\bibitem{Mirzakhani}{M.\ Mirzakhani, ``Simple Geodesics and Weil-Petersson Volumes Of Moduli Spaces of Bordered Riemann Surfaces,'' \emph{Invent.\ Math.\ } \textbf{167} (2007) 179,\\
M.\ Mirzakhani, ``Weil-Petersson Volumes And Intersection Theory On The Moduli Space Of Curves,'' \emph{Journal of the American Mathematical Society} \textbf{20} (2007) 1.}

\bibitem{Wittentopo}{E. Witten, ``On The Structure Of The Topological Phase Of Two-Dimensional Gravity,'' \npb{340}{1990}{281},\\
E. Witten, ``Two-Dimensional Gravity And Intersection Theory On Moduli Space,'' 
\emph{Surveys Diff.\ Geom.\ }\textbf{1} (1991) 243,\\
M. Kontsevich, ``Intersection Theory On The Moduli Space Of Curves And The Matrix Airy Function,'' \cmp{147}{1992}{1}.}

\bibitem{DWreview}{R.~Dijkgraaf and E.~Witten, \emph{Developments In Topological Gravity}, \ijmpa{33}{2018}{30}, arXiv:1804.03275.}

\bibitem{2dgravcont}{F.~David, ``Conformal Field Theories Coupled to 2D Gravity in the Conformal Gauge,'' \mpla{17}{1988}{1651},\\
J.~Distler and H.~Kawai, ``Conformal Field Theory and 2D Quantum Gravity,'' \npb{321}{1989}{509}.}

\bibitem{holomod}{A.~Almheiri and J.~Polchinski, ``Models of AdS$_{2}$ backreaction and holography,'' \jhep{11}{2015}{014}, arXiv:1402.6334,\\
K.~Jensen, ``Chaos in AdS$_2$ Holography,'' \prl{117}{2016}{111601}, arXiv:1605.06098,\\
J.~Maldacena, D.~Stanford and Z.~Yang, ``Conformal symmetry and its breaking in two dimensional Nearly Anti-de-Sitter space,'' \emph{PTEP} {\bf 12} (2016) 12C104 arXiv:1606.01857,\\
J.~Engels\"oy, T.~G.~Mertens and H.~Verlinde,``An investigation of AdS$_{2}$ backreaction and holography,'' \jhep{07}{2016}{139}, arXiv:1606.03438.}

\bibitem{Kitaev}{
  A.~Kitaev,
  ``A Simple Model of Quantum Holography,''
  KITP Program \emph{Entanglement in Strongly-Correlated Quantum Matter}, unpublished,
  see http://online.kitp.ucsb.edu/online/entangled15/,\\
  A.~Kitaev and S.~J.~Suh,
  ``The soft mode in the Sachdev-Ye-Kitaev model and its gravity dual,'' \jhep{05}{2018}{183}, arXiv:1711.08467,\\
  S.~Sachdev,
  ``Holographic metals and the fractionalized Fermi liquid,'' \prl{105}{2010}{151602}, arXiv:1006.3794,\\
S.~Sachdev and J.~Ye,
  ``Gapless Spin-Fluid Ground State in a Random Quantum Heisenberg Magnet,''
  \prl{70}{1993}{3339}, cond-mat/9212030.}
  
\bibitem{tensors}{R.~Gurau,
  ``The $1/N$ expansion of colored tensor models,''
  \emph{Annales Henri Poincar\'e} \textbf{12} (2011) 829, arXiv:1011.2726,\\
 E.~Witten,
  ``An SYK-Like Model Without Disorder,'' \jpa{52}{2019}{47}, arXiv:1610.09758,\\
  I.~R.~Klebanov and G.~Tarnopolsky,
  ``Uncolored random tensors, melon diagrams, and the Sachdev-Ye-Kitaev models,'' \prd{95}{2017}{046004}, arXiv:1611.08915,\\
F.~Ferrari,
  ``The Large $D$ Limit of Planar Diagrams,'' \emph{Ann.\ Inst.\ Henri Poincar\'e Comb.\ Phys.\ Interact.}\ \textbf{D6} (2019) 427, arXiv:1701.01171, 
  F.~Ferrari, V.~Rivasseau and G.~Valette,
  ``A New Large $N$ Expansion for General Matrix-Tensor Models,'' \cmp{370}{2019}{403}, arXiv:1709.07366.}
  
\bibitem{SSS}{P.~Saad, S.~H.~Shenker and D.~Stanford, ``JT gravity as a matrix integral,'' arXiv:1903.11115.}

\bibitem{Eynard}{B.~Eynard, ``Topological expansion for the 1-Hermitian matrix model correlation functions,'' \jhep{11}{2004}{031}, arXiv:hep-th/0407261,\\
B.~Eynard and N.~Orantin, ``Invariants of algebraic curves and topological expansion,'' \emph{Commun.\ Num.\ Theor.\ Phys.\ }\textbf{1} (2007) 347, arXiv:math-ph/0702045,\\
B.~Eynard and N.~Orantin, ``Weil-Petersson volume of moduli spaces, Mirzakhani's recursion and matrix models,'' arXiv:0705.3600.}

\bibitem{pslform}{T.~Fukuyama and K.~Kamimura, ``Gauge Theory of Two-Dimensional Gravity,'' \plb{160}{1985}{259},\\ Phys. Lett. 160B, 259 (1985),\\
K.~Isler and C.~Trugenberger, ``Gauge Theory of Two-Dimensional Quantum Gravity,'' \prl{63}{1989}{834},\\
A.~Chamseddine and D.~Wyler, ``Gauge Theory of Topological Gravity in 1 + 1 Dimensions,'' \plb{228}{1989}{75}.}

\bibitem{Verlinde19}{L.~V.~Iliesiu, S.~S.~Pufu, H.~Verlinde and Y.~Wang,
``An exact quantization of Jackiw-Teitelboim gravity,'' \jhep{11}{2019}{091}, arXiv:1905.02726.}

\bibitem{MertensetalDefects}{T.~G.~Mertens and G.~J.~Turiaci, ``Defects in Jackiw-Teitelboim Quantum Gravity,'' \jhep{08}{2019}{127}, arXiv:1904.05228.}

\bibitem{Wittensummer}{E.~Witten, ``Matrix Models and Deformations of JT Gravity,'' arXiv:2006.13414.}.

\bibitem{Ferlectures}{F.~Ferrari, \emph{Notes on Hyperbolic Quantum gravity}, to appear.}

\bibitem{Goel}{A.~Goel, L.~V.~Iliesiu, J.~Kruthoff and Z.~Yang,
``Classifying boundary conditions in JT gravity: from energy-branes to $\alpha$-branes,''
arXiv:2010.12592.}

\bibitem{Brendle}{B.~Osgood, R.~Phillips, P.~Sarnak, ``Extremals of determinants of Laplacians,'' \emph{J.\ Funct.\ Anal.\ }\textbf{80} (1988) 148,\\
S.\ Brendle, ``A family of curvature flows on surfaces with boundary,'' \emph{Math.\ Z.\ }\textbf{241} (2002) 829.}

\bibitem{Rosenberg}{H.~Rosenberg, ``Constant mean curvature surfaces in homogeneously regular 3-manifolds,'' \emph{Bull.\ Austral.\ Math.\ Soc.\ }\textbf{74} (2006) 227.}

\bibitem{Lopez}{R.~L\'opez-Soriano, A.~Malchiodi and D.~Ruiz, ``Conformal metrics with prescribed Gaussian and geodesic curvatures,'' arXiv:1806.11533 [math.AP].}

\bibitem{Rupflin}{M.~Rupflin,``Hyperbolic Metrics on Surfaces with Boundary,'' \emph{J.\ Geom.\ Anal.} (2020), arXiv:1807.04464 [math.DG].}

\bibitem{Randol}{B.~Randol, ``Cylinders in Riemann surfaces,'' \emph{Comment.\ Math.\ Helv.\ }\textbf{54} (1979) 1.}

\bibitem{Sun}{T.~Sun, ``A note on constant geodesic curvature curves on surfaces,'' \emph{Ann.\ I.\ H.\ Poincar\'e} \textbf{26} (2009) 1569.}

\end{thebibliography}
\end{document}